\RequirePackage{fix-cm}
\documentclass[runningheads,natbib]{svjour3a} 
\usepackage{graphicx}
\usepackage{natbib}
\usepackage{amssymb}
 \journalname{The Astronomy and Astrophysics Review}

\newcommand{\araa}{Ann\ Rev\ Astron\ Ast\-ro\-phys~}

\newcommand{\aapr}{Astron\ Ast\-ro\-phys\ Rev~}
\newcommand{\aap}{Astron\ Astrophys~}
\newcommand{\aaps}{Astron\ Astrophys Suppl~}

\newcommand{\apj}{Ast\-ro\-phys\ J~}
\newcommand{\aj}{Ast\-ro\-nom\ J~}
\newcommand{\apjl}{Ast\-ro\-phys\ J\ Lett~}
\newcommand{\apjs}{Ast\-ro\-phys\ J\ Suppl~}

\newcommand{\jgr}{J\ Geo\-phys\ Res~}
\newcommand{\mnras}{Monthly Not\ Royal Astron\ Soc~}
\newcommand{\pasj}{Publ  Astron\ Soc Japan~}
\newcommand{\nat}{Nature~}
\newcommand{\nar}{New Astronomy Reviews~}
\newcommand{\na}{New Astronomy~}
\newcommand{\sovast}{Soviet Astronomy Lett~}

\newcommand{\prd}{Phys\ Rev\ D~}

\newcommand{\ssr}{Space Sci\,Rev~}

\newcommand{\jcap}{J. Cosmol. Astropart. Phys. ~}

\def\lsim{\;\raise0.3ex\hbox{$<$\kern-0.75em\raise-1.1ex\hbox{$\sim$}}\;}
\def\gsim{\;\raise0.3ex\hbox{$>$\kern-0.75em\raise-1.1ex\hbox{$\sim$}}\;}

\def\cmc{\rm ~cm^{-3}}
\def\em{\rm ~cm^{-6}~pc^3}

\def\cm2{\rm ~cm^{2}}
\def\kms{\rm ~km~s^{-1}}
\def\ergs{\rm ~erg~s^{-1}}
\def\sfr{\rm \Msun  yr^{-1}}
\def\sfrd{\rm \Msun  yr^{-1} kpc^{-2}}
\def\surfmd{\rm \Msun  kpc^{-2}}

\def \kms {\rm ~km~s^{-1}}

\def\ergs{\rm ~erg~s^{-1}}
\def\ergssq{\rm ~erg~s^{-1}~kpc^{-2}}

\def\lfl{\rm ~ph~cm^{-2}~s^{-1} }

\def\enfl{\rm ~erg~cm^{-2}~s^{-1}}
\def\arcmin{\hbox{$^\prime$}}
\def\arcsec{\hbox{$^{\prime\prime}$}}

\def\fl{\rm ~cm^{-2}~s^{-1}}
\def\be{\begin{equation}}
\def\ee{\end{equation}}

\newcommand{\Msun}{\mbox{$\rmm M_{\odot}\;$}}
\newcommand{\msun}{\mbox{$\rmm M_{\odot}\;$}}
\newcommand{\Lsun}{\mbox{$\rmm L_{\odot}\;$}}
\newcommand{\Mdot}{\mbox{$\dot{{\rmm M}}$}}
\newcommand{\Edot}{\mbox{$\dot{{\rmm E}}$}}
\newcommand{\lsun}{\mbox{$\rmm L_{\odot}\;$}}

\def\lsim{\;\raise0.3ex\hbox{$<$\kern-0.75em\raise-1.1ex\hbox{$\sim$}}\;}
\def\gsim{\;\raise0.3ex\hbox{$>$\kern-0.75em\raise-1.1ex\hbox{$\sim$}}\;}

\newcommand{\pcc}{\,cm$^{-3}$}

\newcommand{\rmm}{\mathrm}
\begin{document}

\title{\textsf{Nonthermal particles and photons in starburst regions and superbubbles}}


\titlerunning{Nonthermal phenomena in superbubbles}        

\author{\textsf{Andrei~Bykov}}


\institute{A M Bykov \at
             Ioffe Institute, 194021
St.Petersburg, Russia\\ \email{byk@astro.ioffe.ru}}

\date{Received: date / Accepted: date}

\maketitle
\vspace{-0.3cm}
\begin{abstract}
Starforming factories in galaxies produce compact clusters and
loose associations of young massive stars.
Fast radiation-driven winds and supernovae
input their huge kinetic power into the interstellar
medium in the form of highly supersonic and superalfvenic outflows.
Apart from gas heating, collisionless relaxation of fast plasma outflows
results in fluctuating magnetic fields and energetic particles.
The energetic particles comprise a long-lived component which may
contain a sizeable fraction of the kinetic energy released
by the winds and supernova ejecta and thus modify the magnetohydrodynamic
flows in the systems. We present a concise review of observational
data and models of nonthermal emission from starburst galaxies,
superbubbles, and compact clusters of massive stars.
Efficient mechanisms of particle acceleration and amplification
of fluctuating magnetic fields with a wide dynamical range
in starburst regions are discussed. Sources of cosmic rays, neutrinos
and multi-wavelength nonthermal emission associated with
starburst regions including potential galactic "PeVatrons"
are reviewed in the global galactic ecology context.
\end{abstract}
\maketitle

\section{Introduction}

Almost eighty years ago W.~Baade and F.~Zwicky proposed a very
profound idea that supernovae are the sources of the bulk of
relativistic particles that are detected at the Earth as cosmic rays
(CRs). This idea connects the problem of the origin of cosmic rays to
the problem of stellar evolution -- the crucial phenomenon in physics
of galaxies and clusters of galaxies. This connection is not just a
one way relationship, as  it puts the non-thermal components into the
context of interstellar and intergalactic gas dynamics with tight
links to cosmic magnetic fields, chemical evolution of matter, star
formation (SF) and other phenomena of fundamental importance. Indeed,
the observed energy density of relativistic particles in the Earth
vicinity (about 1.5 eV $\cmc$) is comparable with energy densities of
the other major constituents of the interstellar medium (ISM) in the
Milky Way. The power of CR sources needed to sustain the observed CR
energy density assuming a quasi-steady CR propagation regime and using
the observed secondary to primary CRs ratios and anisotropy can be estimated as about
3$\times$10$^{38} \ergssq$ with the total  power $\gsim 3\times 10^{40} \ergs$ \citep[see, e.g.,][]{ginzburg64,berea90, smp07}.

The cosmic-ray pressure in violent star forming regions may reach a critical value which is analogous  to the well known Eddington limit resulting in large scale outflows in the form of local chimneys or global galactic wind driven by cosmic rays \citep[see e.g.][]{socrates_CR_Edd08,CR_outflows14}. The CR-driven outflows along with the pressure driven winds \citep{chev_clegg85,norman_ikeuchi89,gal_winds_ARAA05} could affect the star formation rate \citep[see e.g.][and the references therein]{bollatto_nat13}.

In this review we briefly outline the basic properties of SF regions
and starburst phenomena, then concentrate on particle acceleration and
magnetic field amplification processes in their connection to SF, and
finally discuss radiation processes and multi-wavelength spectra of
nonthermal emission of SF regions.

\section{Star formation in the Milky Way and other galaxies}

Young massive stars are the main acting agents in the chemical and
dynamical evolution of galaxies and they are the sources of
non-thermal components -- cosmic rays and associated magnetic fields.
These stars are associated with regions of recent SF activity. The SF
is of fundamental importance for galactic ecology. The Milky Way is a
mildly-barred, gas-rich spiral galaxy. The Galactic disk contains
about (2 -- 6)$\times$10$^9 \Msun$ of atomic hydrogen H~I and (1
-- 3)$\times$10$^9 \Msun$ of molecular hydrogen H$_2$ \citep[see,
e.g.,][]{combesARAA91, katya08}. It is characterised by the most
active SF in the Local Group of about 2 $\sfr$. The estimations of the
rate of gas infall from the Local Group are highly uncertain and range
from 0.1 to 1 $\sfr$. The mean SF efficiency per giant molecular cloud
in the solar vicinity is about 5\% \citep[see, e.g.,][for a
review]{2014arXiv1402.6196M}.

\begin{figure}
\includegraphics[width=300pt]{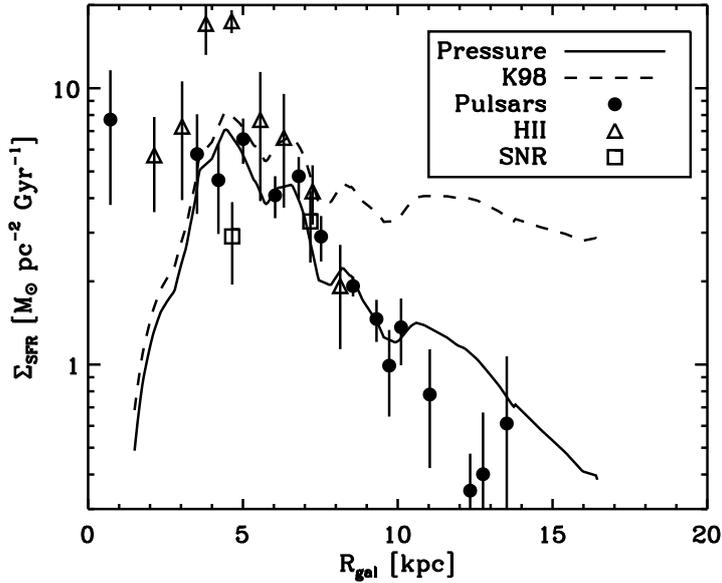}
\caption{Star formation (SF) rate as a function of galactocentric
radius R$_{gal}$ for the Milky Way adopted from Blitz and Rosolowsky
(2006). The SF rate values as derived from observations of H~II
regions by Guesten and Mezger (1982) (triangles), from galactic pulsar
population by Lyne et al. (1985) (circles) and from supernova remnants
Guibert et al. (1978) (squares). The pressure-based SF prescription by
Blitz and Rosolowsky (2006) is given by the solid line to be compared
to that of Kennicutt (1998) (dashed line). The pressure formulation
predicts not only the shape of the curve for the observed SF rate, but
for most of the points, the amplitude as well.} \label{SFR_Blitz06}
\end{figure}

Recent high resolution multi-wavelength observations and numerical
simulations allowed us to look deep into SF though some details of the
process are not very clear as of yet. SF is known to be due to
gravitational collapse of the dense cores of molecular clouds.
Supersonic turbulent gas motions, magnetic fields in the molecular
cloud, as well as radiative and mechanical feedback from the formed
stars each play a role in this process
\citep[e.g.,][]{federatheaApJ11, hennebelleAARv12,
SF_merging_MNRAS14}.

Magnetic fields may provide a support to low
density gas on large scales, while the local gas heating due to
radiation feedback from (proto)stars was found to be able to strongly
suppress  fragmentation on the small scales \cite[see,
e.g.,][]{ineff_SFRMNRAS09}. Both effects were predicted to result in
rather an inefficient SF process in a molecular cloud, with a star
formation rate (SFR) of less than 10\% of the cloud mass for the
free-fall time, that roughly corresponds to the observed rate. On the
other hand, rather surprisingly, current observations support the
empirical star formation laws, which indicate that the area-averaged
SFR depends mostly on the surface density of molecules \citep[see,
e.g.,][]{elmegreen_IAU12}. An analysis of \citet[][]{blitzea06}
revealed that the ratio of atomic to molecular gas $R_{\rm mol}$ in
galaxies is determined by hydrostatic pressure of the interstellar gas
$P_{\rm ext}$. They found that $R_{\rm mol}\propto P_{\rm ext}^{0.92
\pm 0.10}$, at a given radius in a spiral galaxy and that the
relationship between the two holds over 3 orders of magnitude in
pressure. Since SF in giant molecular clouds determine the galactic
SFR, \citet[][]{blitzea06} proposed a pressure-based prescription for
SF in galaxies that is based on the $R_{\rm mol}$--$ P_{\rm ext}$
relation which correctly describes SFR in molecule-poor galaxies
(e.g., in the outer regions of large spirals, dwarf galaxies, damped
Ly${\alpha}$ systems) while for molecule-rich galaxies the
prescription has a form similar to that suggested by
\citet[][]{kennicutt_ApJ98}. In Figure~\ref{SFR_Blitz06} adopted from
\citet{blitzea06}, the SFR in the Milky Way is shown as a function of
galactocentric radius R$_{gal}$. The figure compares the SFR values as
estimated from observations of H~II regions by
\citet{SF_gusten_metzger82}, from galactic pulsar population by
\citet{Lyne_pulsar85}, and from supernova remnants by
\citet{guilbertea78} with results of the pressure-based star formation
prescription by \citet{blitzea06} given by the solid line and that of
\citet[][]{kennicutt_ApJ98}. The pressure formulation not only
predicts the shape of the curve for the SFR observed in the Milky Way,
but also accounts for its amplitude. It may naturally explain the high
SFRs in some galactic merger events. In a recent parsec-resolution
simulations of galaxy mergers \citet{SF_merging_MNRAS14} demonstrated
the triggering of starbursts. They showed that a   gravitational
encounter between two galaxies enhances tidal compression and
compressive turbulent mode thus leading to high SF activity.

The empirical SF law found in galaxy disks suggested an approximately
linear relationship between the SFR per unit area and the molecular
cloud mass per unit area with a timescale for molecular gas conversion
into stars to be about a few Gyr if the gas surface density is $\gsim
10~ \surfmd$ \citep[see, e.g.,][for a
review]{elmegreen11a,elmegreen_11}. On the other hand, the depletion
time, which is usually defined as the ratio of the gas mass to the
SFR, is about 100 Gyr in the outer parts of galaxies and in dwarf
irregular galaxies with the gas surface density below 10$~ \surfmd$
where the disk is dominated by atomic hydrogen. Following
\citet{K_M05} it is instructive to introduce $\epsilon_{\rm ff}$ --
the ratio of the free fall time of self-gravitating gas in molecular
clouds to the gas depletion time. Observations compiled in Figure~4 of
\citet{Krumholz_SFR_PR14} suggested $\epsilon_{\rm ff} \approx 0.01$
as the universal value for galaxies in the local Universe. The complex
process of matter cycling from the ISM gas into stars with the return
of metal rich material with high energy stellar ejecta is the driver
of galactic evolution through cosmic time \citep[see for a recent
review][and references therein]{draine11, 2014arXiv1402.6196M,
Krumholz_SFR_PR14}. Cosmic gas accretion by galaxies affects global
SF. In particular, the relationship between stellar mass, metallicity,
and the SFR (the fundamental metallicity relationship) observed in
large samples of SF galaxies can be explained by metal-poor gas
accretion which is especially important at high redshifts as argued by
\citet{SF_AARv14}. They discussed the global SF history of the
universe and the history of gas around galaxies as inferred from the
observed absorption features in the spectra of background sources.

Synthesis of elements and dust, chemical evolution of galaxies, large
scale motions, cosmic rays, and magnetic fields in ISM are the
integral parts of the global star formation process. While
multi-wavelength observations have yielded a lot of information and
phenomenological models exist, a comprehensive, quantitative theory of
the process is still to be constructed. A challenging issue is the
feedback from newborn clustered stars on the SF environment in
molecular clouds that limits the rate and efficiency of SF.

Observations support general picture of formation of highly
structured, clumped dense molecular clouds from diffuse tenuous ISM
gas resulting in SF with subsequent feedback processes due to
radiation and kinetic energy release from OB associations and compact
clusters of young massive stars. Formation of dense cold clouds in the
ISM is likely a result of multi-fluid gravitational instabilities of
gas shocked by by spiral density waves. The gas motions accelerated by
the radiation and winds of young hot stars clear out cavities in the
intercloud matter and accumulate gas on the periphery of the cavities.

The rate at which the gas is converted into stars is controlled
locally by the radiation of young stars, turbulence, magnetic fields,
and feedback by winds of massive stars and supernovae with the
emergence of superbubbles though large-scale galactic structures such
as the spiral arms and the central bar
\citep[e.g.,][]{ineff_SFRMNRAS09, elmegreen_IAU12,
krumholzea_feedback14, krause14}. The relative role and the links
between the processes listed above are still not very clear, but the
pressure is an important factor. Stellar radiation and evolution of
supernovae at later stages are the sources of gas momentum in SF
regions, which is relaxed in part through shocks into high pressure
gas bubbles and superbubbles. The bubbles may trigger sequential SF
with locally increased SFR \citep{elm_lada77}.

A statistical study of
young stellar objects in the vicinity of 322 mid-infrared bubbles
detected by {\sl Spitzer} \citep{triggeredSFR12} revealed that the
fraction of massive stars in the Milky Way formed by the triggering
process is possibly between 14 and 30\%. Fast shocks of velocity
$v_{\rm sh}$ of a few thousands $\kms$ produce hot gas with
temperature well above 10$^6$ K, which slowly cools thus providing
long lived high pressure bubbles. The shocks are also efficiently
producing nonthermal pressure components consisting from highly
amplified fluctuating magnetic fields and relativistic particles with
partial pressure $P_{\rm nt}$, which may reach a sizeable fraction
$\eta \lsim 0.5$ of the shock ram pressure (i.e., $P_{\rm nt} = \eta
\rho v_{\rm sh}^2/2$). The nonthermal components may affect the
ionization and sources of turbulence in molecular clouds. In this
review we discuss the origins and major properties of the nonthermal
components in SF regions.

The estimated  global SFR in different galaxies ranges by many orders
of magnitude from less than 10$^{-3} ~ \sfr$ in some galaxies with
very inefficient star formation to about 10$^4 ~ \sfr$ in the
so-called ultra luminous infrared galaxies (ULIRGs) which show
far-infrared (FIR) luminosities above 10$^{12}~\lsun$. The average SFR
has a clear trend with the galaxy Hubble type, though a strong
dispersion is present in SFR distribulions of galaxies of the same
type \citet{kennicuttARAA98}. The SFR variations are time dependent
and can be caused by the high dispersion in gas content, galactic
interactions and activity of galactic nuclei.

While the low SFR at the level below 10$^{-2} ~ \sfr$ was revealed by
deep Fabry-Perot H$_{\alpha}$ imaging in some S0 type galaxies, some
other HI-rich S0 galaxies as well as other spirals of types Sa and
later demonstrated somewhat higher SF activity.  Observations with
{\sl Spitzer} and {\sl Herschel} revealed that SFR strongly increases
with redshift for luminous infrared galaxies. Use of  mid-IR
spectroscopy allowed to separate the signal from dust heated by
massive stars from that heated by AGNs.  At low redshifts ULIRGs are
mainly characterized by circumnuclear type of starburst activity
initiated by mergers while the high redshift population of ULIRGs is
dominated by the galaxies with extended SF regions \citet[see,
e.g.,][and references therein]{kennicuttARAA12}. In this review we
mainly focus on active SF regions and starburst phenomena.

\section{Active starforming regions and starburst galaxies}

Possible criteria of the starburst  following are a high SFR with
gas-consumption timescales of less than 1 Gyr and SFR surface density
averaged over the disk above 0.1 $\sfrd $ \citep{kennicuttARAA12}.
The high SFR can not be sustained for longer than a small fraction
(e.g.,  less than 10\%) of the Hubble time. Starburst regions with SFR
of about 100 $\sfr$ are not active for  much longer than 100 Myr. The
typical observational appearances of the starburst in either nucleus
or active regions of a galaxy are (i) the large ratio of the infrared
(IR) to the bolometric luminosity, which is due to re-radiation of the
luminous emission of hot young stars absorbed by the surrounding dust
in the IR band, (ii) high luminosity of Balmer-line emission with
large equivalent widths, or (iii) strong radio continuum emission.

A number of starburst galaxies are already well known, the most studied
among these being the nearby Irr2 type galaxy Messier 82, NGC 253 (a
barred Sc galaxy), and Arp 220 (likely a merger).  Starbursts are
usual phenomena in interacting and merging galaxies. The Antennae
Galaxies are a pair of interacting starburst galaxies  NGC 4038/NGC
4039 located in the Corvus constellation. The observed ongoing phase
of starburst in NGC 4038/NGC 4039 \citep[see
e.g.][]{Whitmore95,Whitmore10} is supporting the suggestions that the
dense high pressure regions in merging galaxies are the regions with
high SFR \citep[see  e.g.][]{blitzea06,SF_merging_MNRAS14}.

The observed FIR --- radio correlation in SF galaxies indicates that
massive stars are the energy source for both FIR and nonthermal radio
emission \citep[see e.g.][]{FIR_radio85, helou_FIR_radio85, voelk89,
condon92, SFRbook07, FIR_radio13, zweibel_IR_radio13}. FIR in
starforming galaxies is mostly produced by the dust heated by the
radiation of stars more massive than 5$\Msun$, and ${\rm L_{60 \mu
m}}$ characteristic luminosity at 60 $\mu m$ is connected to the SFR
as
\begin{equation}
SFR(M  >5\Msun)=  \frac{\rm L_{60 \mu m}}{5.1 10^{23} W Hz^{-1}}\, {\sfr},
\end{equation}\label{sfr_rate}
where  $SFR(M >5\Msun)$ is the rate of the formation of stars heavier
than 5 $\Msun$ \citep[see, e.g.,][]{SFR_cram98}.

Starburst galaxies are known to have high supernova rates in the
active SF regions. Radio and IR observations of M82
\citep[see][]{unger84,kronberg85,voelk89,condon92,huang94,M82_book_07}
revealed many supernova remnants (SNRs). The radio luminosity of an SF
regions is assumed to be proportional to the recent SF rate and is
thought to be connected to  $SFR(M>5\Msun)$.  With a particular model
of the origin of thermal and nonthermal radio continuum radiation it
is possible to disentangle the nonthermal radio luminosity
\citep[e.g.,][]{Biermann76, condon92,SFR_cram98} and thus obtain the
SFR as
\begin{equation}
SFR(M  >5\Msun)=  \frac{\rm L_{1.4 GHz}}{4.0 10^{20} W Hz^{-1}}\, {\sfr}.
\end{equation}\label{sfr_rate1}

\subsection{Initial mass functions of clusters and stars}

The star formation activity results in some hierarchical structure
with both spatial and temporal correlations \citep[see,
e.g.,][]{elmegreen11b}, which is important for ISM ecology and
evolution \citep{McCray87, maclow_mccray88, norman_ikeuchi89,
heiles90, heiles01}. Young massive SF regions are often clumped
\cite{lada_2003ARA&A, massive_star_cluster_2010ARA&A}. Active SF
regions may consist of a number of clusters and extended populations
of massive stars. On the other hand, some of  the large SF regions are
loose OB star associations like Cyg OB2, where no signatures of dense
star clusters was found \citep[see e.g.][]{cygOB2_mnras14}. It was
argued by \citet{Tutukov78,lada_2003ARA&A} that about 90\% of the
clusters disperse within about 10 Myr.

A fundamental question is why
and how the stars that are initially born in high density,
highly-clustered SF regions eventually form a field population, where
the majority of the stars are not bound to any structure smaller than
the galaxy as a whole, while some minor population still resides in
gravitationally-bound clusters with an initial mass function (IMF) at
birth
$dN/dM \propto M^{-\beta}\, \exp(-M/M_{\ast}) $ with $\beta \approx$
2 \citep[e.g.][]{Krumholz_SFR_PR14}.
Observations of young compact clusters indicate that the
characteristic mass $M_{\ast}$ is about 2$\times$10$^5 \Msun$ in the
Milky Way type galaxies, and it is above 10$^6 \Msun$ in luminous
infrared starburst galaxies. It was suggested that young massive
clusters of mass well below 10$^6 \Msun$ may dissolve eventually into
unbound field star populations due to their initial gas expulsion and
some feedback effects. The most massive young clusters similar to
those found in the Antennae galaxies have expected lifetimes of about
the Hubble time and may evolve into old massive clusters, which are
observed in the Milky Way and other galaxies as globular clusters
\citep[see][and the references therein]
{massive_star_cluster_2010ARA&A}.

The present day mass functions of the globular clusters are different
from the IMF of young stellar clusters discussed above. They are peaked
or bell-shaped (for logarithmic mass bins) with the characteristic
mass at about 2$\times$10$^5 \Msun$ which may be either due to
evolution effects that affect the IMF or to the initial form of IMF
\citep[see][for a discussion]{massive_star_cluster_2010ARA&A}. For a
long time globular clusters have been used as natural laboratories for
studies of stellar evolution, relaxation of bound N-body systems, and
stellar encounters which are frequent in the dense central regions of
the globular clusters and may result in formation of a rich sample of
exotic objects of various nature \citep[see, e.g.,][]{meylan97}.

Recent investigations revealed a number of new features of these
objects. It was established that, contrary to a long-standing paradigm
stating that the constituent stars of a globular cluster are born at
the same time and made of the same material, it has recently become
clear that different stellar populations are present there \citep[see
e.g.][]{Tuc47_mixed_pop09}. At least two different stellar populations
were distinguished in 47~Tuc (NGC~104) by \citet{Li_Tuc47_AA14} with
small variation in the iron-peak and neutron capture elements but
different abundances of proton-capture and alpha-elements. This
suggests that a population of the cluster stars exists, which has most
likely been formed from the gas enriched by the stars of the previous
generation.

The form of the star IMF is of fundamental importance for studies of
galaxy evolution. Observations are consistent with the power law
$dN/dM \propto M^{-\alpha} $ where \citet[][]{salpeter_IMF_ApJ55}
deduced $\alpha \approx$ 2.35 valid  for 0.4 $\lsim M/\msun \lsim$ 10.
\citet{kroupa01} proposed a three power-law IMF representation for
Galactic field stars IMF (which is in average populated by 5-Gyr-old
stars) with two characteristic mass values at 0.08 and 0.5 $\msun$.
Namely, for the low mass end 0.08 $\lsim M/\msun \lsim$ 0.5  the IMF
index is $\alpha =$ 1.3, at $M/\msun \leq$ 0.08, it is flatter with
$\alpha =$ 0.3, while in Salpeter's regime $M/\msun \geq$ 0.5 the IMF
index  is $\alpha =$ 2.3. Note, that if $\alpha =$2.0 then equal
masses would be present in every star mass decade.

The problem of universality of the IMF was examined with numerous
observational studies and models \citep[see,
e.g.,][]{salpeter_IMF_ApJ55, scalo_IMF_86, Massey_IMF_1998,
chabrier_baraffe_ARAA00, kroupa_IMF_sci02,
bonnell_ea_IMF_ProtoS-P07,andrewsea13}. A critical review of both the
resolved and the integrated properties of stellar populations in
galaxies by \citet[][]{bastian_IMF_ARAA10} concluded that the IMF is
consistent with the gross of the available data and suggested no
noticeable variations of the IMF over much of cosmic time. The
apparent universality of the stellar IMF is non trivial given the
strong variance of the conditions in SF regions in the galaxies.

Star formation models are connecting the origin of the low-mass stars and
brown dwarfs to fragmentation of dense filaments and disks, while the
continued accretion in a clustered environment limited by the
radiation pressure is the likely mechanism for the high mass SF
\citep[see, e.g.,][and references therein]{bonnell_ea_IMF_ProtoS-P07}.
The maximal stellar mass in the IMF is likely governed by the feedback
processes such as radiation pressure and ionization, though some other
factors may be important as well. Observational estimations range from
150 to 300 $\msun$ with large uncertainties due to unresolved
multiplicity, NIR magnitude-mass relation and some other factors
\citep[see, e.g.,][for a recent
discussion]{massive_star_tan_ProtSP14}. The most massive star reported
by now is R136a1 with an estimated mass substantially exceeding 150
$\msun$ \citep{Crowtherea10}. This Wolf--Rayet star resides in a super
star cluster R136 located in the 30 Doradus region of the Large
Magellanic Cloud (LMC).

The issue of a possible dependence of the IMF upper end of star mass
$m_{max}$ on the total stellar birth mass of the embedded star cluster
$M_{ecl}$ is under discussion \citep[see, e.g.,][]{weidnerea10,
andrewsea13, mmax_mcl_weidner14}. If there is no dependence of the
maximum star mass on the cluster mass,  i.e., the IMF is
stochastically populated, then one can expect that a 10$^5 \msun$
cluster should contain -- in statistical sense -- the stars of the
same masses as a hundred of 10$^3 \msun$ clusters. However, if there
is a positive correlation between the maximal stellar mass and the
cluster mass as proposed by \citet{weidnerea10} then no stars heavier
then about 35 $\msun$ are expected in clusters of 10$^3 \msun$. The
selection criteria used by \citet{weidnerea10} to claim the presence
of the $m_{max}-M_{ecl}$ correlation is based on counting of the
resolved very young stellar populations of age younger than 4 Myr with
no SNRs in the cluster. It was also suggested by
\citet[][]{top_heavyMNRAS11} that in the extreme starburst galaxies
with SFR above 100 $\sfr$ star clusters with masses 10$^6 \Msun$ may
have the top-heavy stellar IMF with the slopes of high mass star
distributions flatter than the standard index of 2.35 observed in the
young massive clusters of mass below 2$\times$10$^5 ~\Msun$.  However,
 \citet{andrewsea13} used multi-wavelength {\sl Hubble Space
Telescope} WFC3 camera data to measure the production rate of the
ionizing photons in unresolved stellar clusters of age below 8~Myr in
the nearby irregular galaxy NGC 4214. With this approach they did not
find any dependence of the maximum stellar mass on the cluster mass
and concluded  that within the uncertainties the upper end of the
stellar IMF appears to be universal in NGC 4214, thus requiring more
observational studies of the existence of the $m_{max}-M_{ecl}$
correlation.

\subsection{Supernovae in starburst regions}

Assuming that all stars which are more massive than 6.7 $\Msun$ become
radio emitting supernovae \citet{condon92} estimated the supernova
rate in SF regions as
\begin{equation}
\nu_{\rm SN} = 0.041 \frac {SFR(M >5\Msun)}{\sfr}~~{\rm yr^{-1}}.
\end{equation}\label{sn_rate}

From observations of radio SNRs \citet{huang94} estimated the suprnova
rate in M82 as 0.11$\pm$ 0.05 ${\rm yr^{-1}}$. At the same time,
\citet{smith_Arp220_snrate98} resolved the radio nucleus of Arp 220
and found many radio-emitting point sources, which were identified as
SNRs. They estimated the SFR in Arp 220 as about 50--100 $\sfr$ and
estimated  he supernova rate  in the galaxy as 1.75-3.5 ${\rm
yr^{-1}}$. Another nearby nuclear starburst galaxy NGC 253 is a barred
Sc spiral galaxy in the Sculptor group located at a distance of about
3.5 Mpc where one arcsecond corresponds to about 17 parsecs.  Recent
radio observations at 2.3 GHz performed with the {\sl Australian Long
Baseline Array} in a combination with  the {\sl Jansky Very Large
Array} were used by  \citet{snrate_ngc253_AJ14} for multi-epoch
studies of  SNRs in NGC 253. Note that to estimate the value of $SFR(M
 >5\Msun)$ with Eqs. (\ref{sfr_rate})---(\ref{sn_rate}) it is enough
to measure just a single parameter. This helps to constrain the
supernova rate in a number of particular cases.

It is very important to find independent ways to estimate both the
SFRs and the supernova rates. Hubble Space Telescope NICMOS [Fe II]
1.26 $\mu$m line detection was proposed by \cite{FeII_sn_rate03} to
independently measure supernova rate in M82. The authors suggested
that the [Fe II] lines may be produced by an SNR population of ages up
10$^4$ years, while radio SNRs are a younger population of a few
hundred years old. \citet{sne_rate_starburst_AA14} found a nearly
linear relationship between the 1.26 $\mu$m [Fe II] luminosity and the
supernova rate in a sample of 11 nearby starburst galaxy centers. This
relationship is valid for normal SF galaxies but does not hold for
extreme ultraluminous galaxies. The authors pointed out that a
somewhat excessive supernova rate measured in NGC 6240 and Arp 220
compared to radio supernova rate estimatio ns is likely caused by
merger shocks that are overexciting the 1.26 $\mu$m [Fe II] emission
line in addition to the excitations produced by SNRs themselves.

Multi-wavelength observations reviewed  by \citet{SFRbook07} are
consistent with the presence of a central SF region in M82 of  about
0.7 kpc scale size with $SFR(M >5\Msun) \sim 2 \sfr$ and the total SFR
of about 10 $\sfr$ within the central 0.3 kpc,  which is 3 times
higher than the SFR of the whole Milky Way. The ultraluminous merger
Arp 220 (IC 4553/4)  with IR luminosity of about 10$^{12.2}\, \Lsun$
in the 8--1,000 $\mu$m band  is located at about 75 Mpc from the Sun.
Since 1$\arcsec$ roughly corresponds to 361 pc at the estimated
distance of Arp 220, very long baseline interferometry images at  2 cm
and 3.6 cm wavelength were used to study compact radio continuum
sources in this galaxy \citep[see, e.g.,][]{Batejat11}. The authors
resolved a number of SNRs of diameters above 0.27 pc, whose radio
luminosities are consistent with the standard diameter --- surface
brightness relation. The ratio of magnetic field energy density to
that of relativistic particles was estimated to be about 1\% and
magnetic field strengths were found to be about 15--50 $\mu$G  in the
SNR shells, while an upper limit of 7 $\mu$G was suggested for the ISM
magnetic field in Arp 220. Submillimeter Array imaging of Arp 220 by
\citet[][]{sakamoto_Arp220_submApj08} in the 860 $\mu$m continuum and
CO emission line reached a resolution of 0.23$\arcsec$ (80 pc) and
0.33$\arcsec$ (120 pc), respectively. The submillimeter continuum
emission dominated by emission of heated dust morphologically agrees
with the distribution of radio supernovae in the eastern nucleus,
while in the bright western nucleus the dust emission is more compact
than the supernova distribution. This possibly indicates that the dust
heating in the western part of the galaxy is mainly due to the
activity of the nucleus (AGN), though a presence of super star cluster
can not be excluded yet. AGNs themselves may seriously affect the SF
phenomena.

High resolution near IR adaptive optics imaging of the
central 100 pc region in  the Seyfert I NGC~2110 galaxy with the {\sl
Keck OSIRIS} instrument \citep{starb_cl_ngc2110} revealed a number of
star clusters in the central 90$\times$ 35 parsec bar where [Fe II]
emission indicated the shocked material. Each of these clusters is a
few times brighter than the Arches cluster in the Galactic Center
region of the Milky Way. The SFR in the central core of NGC~2110 is
about 0.3 $\sfr$. If most of the energy input from supernova and
stellar winds in the localized SF regions is thermalized, a strong
wind can be driven out of an active region.  \citet{chev_clegg85}
presented a very useful analytical solution for the wind which is
driven from a region of a uniform mass and energy deposition. The
solution is parameterized by the uniform mass and energy input rates
within a SF region of a given radius $R$, which was assumed to be
about 200 pc for a starburst galaxy M82. The non-thermal components
including magnetic fields and relativistic particles which we shall
discuss below may affect the wind properties in a number of aspects.

\begin{figure}
\includegraphics[width=300pt]{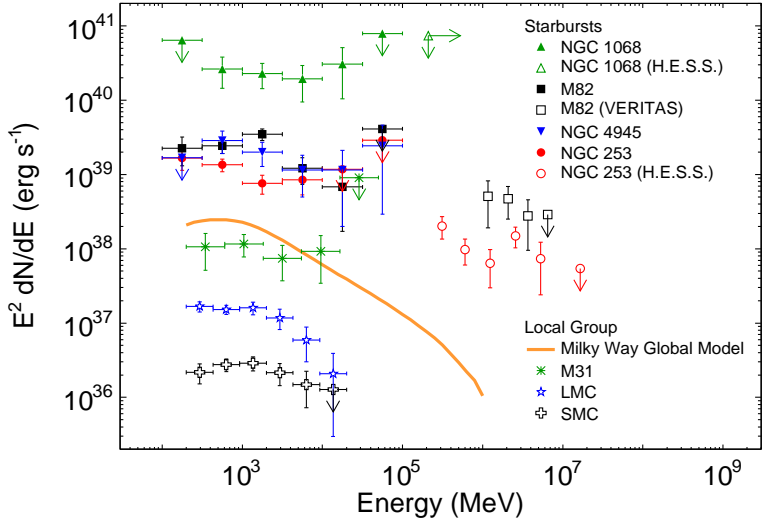}
\caption{The gamma-ray spectra of starburst galaxies M82, NGC 253, NGC
1068 and others detected by the Large Area Telescope aboard {\sl
Fermi} observatory and imaging atmospheric Cherenkov telescopes adopted
from \citet[][]{Fermi_SF_ApJ12}} \label{fig:SFG_Fermi_TeV}
\end{figure}

TeV range gamma-ray photons were detected from NGC~253 with the
H.E.S.S. ground-based Cherenkov imaging atmospheric telescope
\citep[see, e.g.,][]{acero_SFgalaxies_Sci09, HESS_NGC253_ApJ12}. GeV
regime emission was detected with the {\sl Fermi Large Area Telescope}
from the starburst galaxies M82, NGC 253, NGC 1068, NGC 4945
\citep[see, e.g.,][]{Fermi_M82_NGC253_ApJ10,Fermi_SF_ApJ12} with
gamma-ray luminosity ranging from 10$^{39}$ to 10$^{41}\,\ergs$. In
Fig.~\ref{fig:SFG_Fermi_TeV} adopted from \citet{Fermi_SF_ApJ12}
gamma-ray luminosities of starburst galaxies are shown in comparison
with the Local Group sources. Depending on the particular mechanism of
emission of the accelerated particles the measured gamma-ray
luminosity may constrain the energy content in the non-thermal
components, i.e., relativistic particles and magnetic fields
\citep[see, e.g.,][]{Ohm_gamma_starburst12, Torres_Reimer13, Lacki13,
martin_14_star_form_gamma,yoast_hull_NGC253_1068}. For instance,  \citet[][]{Lacki11} argued
that if the gamma-rays observed from M82 and NGC 253 are of hadronic
origin, i.e., they come from pion decays following inelastic
collisions of relativistic nuclei with the interstellar matter, then
20--40\% of the energy injected in the high-energy primary CR protons
is lost in the inelastic collisions before escape from the starburst
region. On the other hand, \citet{mannheimea_PWN_SF_APh12} argued that
apart from the hadronic emission from CRs the gamma-ray pulsar wind
nebulae left behind by the supernovae  may substantially contribute to
the observed TeV luminosity of starburst galaxies. They pointed out
that the hard emission detected from NGC~253 can be attributed to
about 3$\times 10^4$ pulsar wind nebulae expected in a typical
starburst galaxy.

The starburst  galaxies can contribute significantly to the diffuse gamma-ray background.  Recent modeling by \citet{starburst_neutrinos14}  suggested that starbursts, including those with active galactic nuclei and galaxy mergers, could be the main sources of the high-energy neutrinos observed by the {\sl IceCube Observatory} (see also \citet{2014PhRvD..89l7304A}). Assuming a cosmic-ray spectral index of the diffuse CR population to be 2.1--2.2 for all starburst-like galaxies and extrapolating from GeV to PeV energies the authors obtained fluxes consistent with both the Fermi and IceCube data. This extrapolation assumes that there is no spectral break at PeV regime as it is the case in the Milky Way. To justify this one have to find a population of Pevatrons in the starburst galaxies. We shall argue in  \S\ref{CSF} that supernova shocks colliding with
the fast winds in the compact massive star (super)clusters like  Westerlund I (see \S\ref{Westerlund}) can accelerate CRs well above PeV. Since the superclusters are likely to be much more frequent in starburst galaxies CRs escaped from these accelerators  may support a diffuse CR population  with spectral slope harder than that observed in the Milky Way and consistent with that derived by \citet{starburst_neutrinos14}. Other possible PeV neutrino sources were proposed recently including  semirelativistic hypernova remnants by \citet{2014PhRvD..89h3004L} and shock acceleration of PeV CR particles in massive galaxy mergers or collisions by \citet{CRs_mergers_ApJ14}. In the case of  massive galaxy mergers/collisions CR acceleration in colliding shock flows modeled by \citet{MNRAS_BGO13} may play a role.

\subsection{Galactic Center clusters}

Young massive star clusters of M  $>$ 10$^5 \Msun$ typically contain
hundreds of thousands to millions of young stars within a parsec scale
size core. They are called ``super star clusters'' (or superclusters
hereafter) and are found in starburst galaxies with high SF rate
\citep[see, e.g.,][]{massive_star_cluster_2010ARA&A}. The formation of
these large and dense superclusters was preferebly associated by
\cite{starcluster10} with irregular and interacting galaxies. The lack
of rotational support in giant molecular clouds highly increases the
probablitity of supercluster formation contrary to the picture in
spiral disk galaxies like Milky Way where only a few superclusters
have been found.

There is a number of impressive examples of such
objects in the Local Group including the R136a cluster in Doradus 30
region of LMC, which contains as many as 39 O3 class stars -- the
hottest and most luminous massive stars resolved within spectroscopic
observations of  the {\sl Hubble Space Telescope} by
\citet{massey_R136_ApJ98} with the total estimated stellar mass of
about (2-3) $\times 10^4~\msun$. \citet{massey_R136_ApJ98} also
established that the number of high-mass stars is consistent with that
predicted by the Salpeter-like IMF if it holds over the mass range
2.8--120 $\msun$. However, in a somewhat similar Galactic system
NGC~3603 --- a massive star-forming H II region with the total mass of
the associated molecular cloud about 4$\times 10^5~\msun$ where the
starburst cluster of the estimated stellar mass (1-1.6) $\times
10^4~\msun$ consists of six O3 class stars, three WNL stars,  and many
late O- and B-class stars --- \citet{NGC3603_ApJ08} found that IMF can
be well fitted with a power law of index $\alpha \sim$  1.74 within
the mass range of 0.4 to 20\,$\msun$. Such an IMF fit is substantially
flatter then Salpeter's IMF.

Due to the extremely heavy extinction by the dust in the inner Galaxy
stellar clusters are obscured in the optical band. However, radio,
infrared, and X-ray observations revealed a number of young massive
stellar clusters in the inner Galaxy. Within the last twenty years the
Arches and the Quintuplet clusters, each of mass about 2$\times
10^4~\msun$, were discovered in the 100 pc vicinity of the Galactic
Center and thoroughly studied \citep[see, e.g.,][and references
therein]{Figer08_massive_clusters}.  The SF rate within the central
$\sim$ 200 pc of the Milky Way averaged over $\sim$ 10~Gyr was
estimated to be within 0.04-0.12 $\sfr$
\citep{Crocker_NT_GC_MNRAS12}, though local starburst events are very
likely to occur in some epochs.
The age of the Arches cluster is estimated as 3.5 $\pm$  0.7 Myr and
the Quintuplet cluster as  4.8 $\pm$ 1.1 Myr
\citep[see][]{Schneiderea_ages_Arches_Quint14} thus supporting the
idea of a number of local starburst events in the Galactic Center
neighbourhood during the last 10 Myrs. The  ionizing photon luminosity
of the Arches cluster is about  4$\times$10$^{51}$ photons s$^{-1}$
with the total luminosity of about 6$\times$ 10$^7~\Lsun$ and its
stellar IMF appears to be of "top-heavy" type  for the cluster with
the estimated index $\alpha \sim$  1.6 $\pm$ 0.1
\citep{Arches_clusterApJ02}, which is similar to that of NGC 3603. The
Arches cluster is very compact, its radius is about 0.2~pc providing a
very high mass density 3$\times$10$^{5}~\msun {\rm pc}^{-3}$, while
the size of the Quintuplet star cluster is about a parsec.

Thermal-like and nonthermal X-ray emission was detected from both the
Arches and Quintuplet star clusters with {\sl Chandra} by
\citet{Arches_Quint_clusters_ApJ04} who found a few point like sources
and sources of diffuse thermal emission near the cores of the
clusters. The radius of the core of the Arches cluster is about
5\arcsec thus requiring the use of {\sl Chandra}  to resolve the
point-like source contribution. A number of radio sources with  X-ray
counterparts were studied in both clusters with the {\sl Very Large
Array}. Some of the radio sources have been interpreted to be
originating from the shocked colliding winds of massive binary stars
\citet{VLA_Arches_Quint_ApJ05}. The authors interpreted three
unresolved very compact sources of $<$ 0.01 pc size in the Quintuplet
to be shocked stellar winds and estimated the total stellar wind mass
injection rate for the cluster to be $\Mdot \sim 7 \times
10^{-4}~\sfr$ while it is $ \sim 3\times 10^{-4}~\sfr$ for the Arches
cluster. The mass-loss rates estimated from the radio data imply large
kinetic power of cluster winds to be about $\gsim 10^{38} \ergs$. Such
winds were considered to be the reason of the parsec scale size
emission zones observed in the clusters with diffuse X-ray luminosity
somewhat above 10$^{34}~\ergs$ \citep[see
e.g.][]{rockefellerea_X_arches_quint_ApJ05}.

Apart from the thermal
diffuse emission coming from optically thin plasma at temperatures
$\sim$ 2 keV, non-thermal X-ray emission with relatively hard photon
spectral index $\Gamma \sim$ 1.4 and prominent Fe 6.4 keV K-shell line
emission was revealed with a 100 ks long {\sl Chandra} observation by
\citep{wang_Arches_MNRAS06} at the sub-arcminute large South-Eastern
extension of the Arches cluster. The lightcurves of the Fe K-shell
line emission from three bright molecular knots around the Arches
cluster were found  by \citet{Arches_XMM_AA11} to be constant over 8
years of {\sl XMM-Newton} observations.
\citet{CR_Arches_Tatischeff_AA12} studied the bright 6.4 keV Fe line
features and the non-thermal emission around the Arches cluster with
{\sl XMM-Newton}. They found the metallicity of the ambient plasma in
the cluster to be 1.7 $\pm$ 0.2 times the solar value and attributed
the X-ray line emission to cosmic ray protons hitting the surrounding
molecular gas. These authors pointed out that the Arches cluster
region may be a GeV regime gamma-ray source potentially detectable
with the {\sl Fermi Gamma-ray Space Telescope} and may also produce
gamma-ray line emission \citep[see
also][]{dogiel11,dogiel_GC_lines_11}.  The model of
\citet{CR_Arches_Tatischeff_AA12} requires large fluxes of low-energy
cosmic ray ions, which can be produced if the ions are accelerated
with the efficiency of a few percent at the shocks associated with the
massive star cluster and some adjacent molecular cloud. Indeed such an
efficiency was found plausible in a non-linear model of CR
acceleration by shocks of velocities $\sim$ 1,000 $\kms$ in
superbubbles produced by OB stars \citep[see][]{b01}. Analyzing the Fe
K-shell line emission from the G0.13-0.13 molecular cloud in the
Galactic Center region altogether with the 74 MHz radio continuum
emission \citet{YZ_GC_electron_13} suggested that the Fe line
production is due to the low-energy CR electrons interacting with the
ambient matter.

\begin{figure}
\includegraphics[width=130pt]{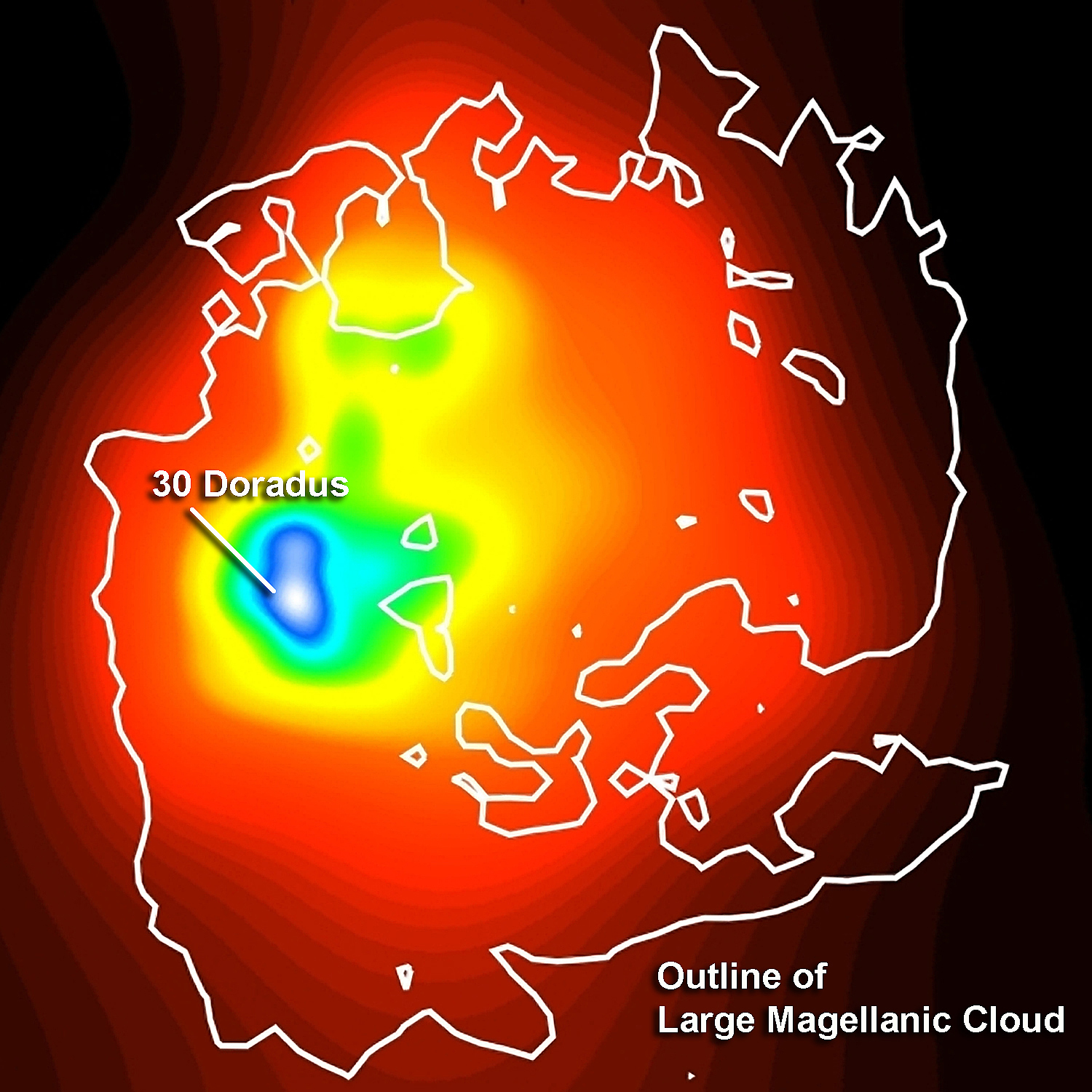}
\includegraphics[width=200pt]{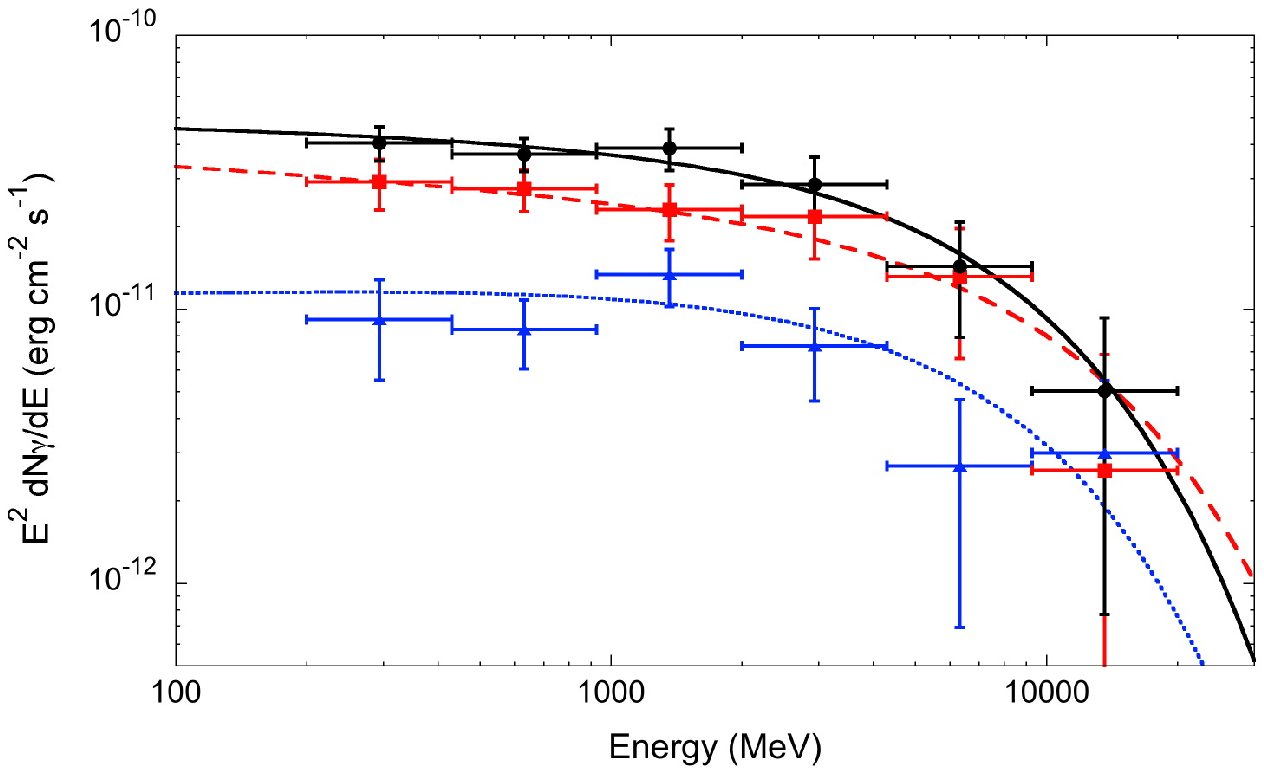}
\caption{{\it Left Panel:}The gamma-ray image of the Large Magellanic
Cloud obtained with the  Large Area Telescope aboard the {\sl Fermi
Gamma-Ray Space Telescope}. Bright colored region shows giant
star-forming region 30~Doradus. Credit: NASA/DOE/Fermi LAT
Collaboration.  {\it Right Panel:} Spectrum of the gamma-ray emission
from the Large Magellanic Cloud in the energy range 200 MeV -- 20 GeV.
Blue dots show the spectrum of 30~Doradus, while black and red dots
show  the total spectrum obtained by fitting the HII gas map and for
the LMC disk, respectively. Error bars include statistical and
systematic uncertainties. Adopted from \citet[][]{LMC_Fermi_AA10}.}
\label{fig:30Dor_Fermi}
\end{figure}

The origin of the Fe K-shell line emission in the Galactic Center
sources is a very interesting and still open problem.  Being
illuminated by powerful hard X-ray radiation from am outburst of the
supermassive black hole at the Galactic Centre the neutral matter of
molecular clouds can reflect and reprocess the radiation into the
observed hard continuum and fluorescent 6.4 keV line emission
\citep{Sunyaevea_ApJ93,SC_MNRAS98}. Indeed, \citet{revnivtsev_GC04}
associated the hard X-ray source IGR~J17475-2822 discovered with the
{\sl International Gamma-Ray Astrophysics Laboratory, INTEGRAL} with
the giant molecular cloud Sgr~B2 in the Galactic Center vicinity. They
argued that the broadband spectrum of X-ray emission of Sgr~B2 is due
to Compton scattered and reprocessed  radiation emitted by a flare of
Sgr A$^{\star}$ 300--400 years ago and fitted the data with a model of
the central black hole, whose 2--200 keV luminosity is about
1.5$\times 10^{39} \ergs$ and the photon spectral index is $\Gamma
\approx$ 1.8. Later \citet{terrier_echo_ApJ10} analyzed 20~Ms of {\sl
INTEGRAL} exposure of the Galactic Center region and found a
significant weakening of the hard X-ray emission from the source in
Sgr~B2 associated with a molecular cloud, with characteristic time of
8.2 $\pm$ 1.7 yrs. This fact supports the interpretation of the Sgr~B2
source as an echo of a past activity of the central black hole, which
came to an end only 75 -- 155 years ago.  In the current quiet state
the central black hole of mass about 4$\times 10^6~\msun$ shows
luminosity less that 10$^{-10}$ of its Eddington limit.

The extended characteristic 6.4 keV line emission from low charge states of Fe
being naturally produced by fluorescent scattering of the X-rays may
be also excited by the low energy cosmic rays penetrating into
molecular clouds in the vicinity of particle acceleration sites
\citep[see, e.g.,][]{valiniaea00,dogiel11, CR_Arches_Tatischeff_AA12}.
Fast moving metal rich fragments of supernova ejecta may be  sources
of both 6.4 keV Fe line emission as well as the high charge state Fe
K-shell lines in 6.7-6.9 keV range \citep{b02,b03,bocchinoea12}.
Cataclysmic variables and coronally active binaries comprising a
numerous population in the Galactic ridge are likely X-ray sources
with prominent Fe lines and were proposed by
\citet[][]{revnivtsevea09} to contribute substantially to the
Galactic Center X-ray emission observed by \citet{munoea09} in deep
{\sl Chandra} exposure.

The Arches cluster was recently observed above 10 keV with {\sl
NuSTAR} \citep{Arches_NuStar14} in an attempt to establish the nature
of the extended non-thermal X-ray emission and Fe line sources. It was
found that the spectrum of the hard X-ray emission extends at least up
to about 40 keV and it was not attributed to any point-like source in
the Arches core, but rather shown to come from a truly extended
region. The  spectrum revealed by  {\sl  NuSTAR}  from a vicinity of
the Arches cluster  is consistent with that of thin thermal plasma of
a temperature about 1.7~keV which was previously found in the cluster
core region and attributed to cluster winds thermalisation, but also
contains a power-law component of photon index $\sim$ 1.6. An
interpretation of the hard X-ray {\sl  NuSTAR} observations can be
suggested both within the context of the low-energy CRs interacting
with the ambient cloud and within the X-ray reflection model.  The
X-ray reflection model is plausible if the X-ray illuminating source
is associated with Sgr A$^{\star}$ flare activity in the past, but not
with the past activity of the Arches cluster itself. For the CR model
the authors estimated the power of the CR proton injection above 10
MeV nucleon$^{-1}$ in the cluster vicinity to be about (5--8)$\times
10^{38} \ergs$. Given the estimated above total stellar wind energy
injection rate and the lack of apparent signatures of SNRs, such an
injection would require some highly efficient process of kinetic power
conversion into accelerated CRs. Strong ultraviolet radiation and
stellar winds must have cleared and strongly heated the gas and dust
in the compact clusters and thus the observational appearance of SNRs
in the clusters may differ from that typical for SNRs in other phases
of the ISM.

\subsection{Westerlund I supercluster and the magnetar CXOU~J1647-45}\label{Westerlund}

The cluster Westerlund 1 with estimated mass $\gsim 5\times10^4 \msun$
is more massive than Arches and Quintuplet,
but its radius is $\sim$ 1 pc resulting in the mass density $\sim 1.2
\times 10^4\, \msun$pc$^{-3}$ which is somewhat below that in the
Arches cluster \citep[see, e.g.,][]{Figer08_massive_clusters,
Clarkea10}. The cluster of the estimated age of 5.0 $\pm$ 1.0 Myr is
located at a distance of about 3.8~kpc (where 1$\arcmin$  is about 1
pc)  in the constellation of Ara. The derived IMF slope $\alpha = 1.8
\pm 0.1$ is of 'top heavy' type,  i.e., it is flatter than that of
Salpeter's IMF \citep[see][and  refernces
therein]{Westerlund1_Lim_AJ13}.  \citet{munoea06} argued that for IMF
with slope 1.8 $\leq \alpha  \leq$ 2.7 for M$\geq$ 30$\msun$ the
cluster should have originally contained from 80 to 150 stars of masses
above 50 $\msun$, which must have exploded as supernovae within the
estimated age of the cluster.  Then the average supernova rate over the
last 1 Myr should be one in about 0.01 Myrs.

The supernovae would provide a very strong source of kinetic power of about 3$\times
10^{39} \ergs$ over a few Myrs time period. Observations indicated
that if the scenario is appropriate only a small fraction $\lsim$
10$^{-5}$ of the kinetic power estimated above is radiated in X-rays
in the 5$\arcmin$ vicinity of Westerlund 1. However, an ejecta of a
supernova exploding in the cluster central region would pass through
the cluster filled with tenuous hot plasma sooner than in 1,000 yrs,
and thus most of the time the cluster would be free from supernova
ejecta.

Diffuse X-ray emission from Westerlund~1 was revealed with {\sl
Chandra} observations by \citet{munoea06} who was able to subtract the
point-like X-ray source contamination from the observed emission of
the cluster down to a completeness limit of $\sim$
2$\times$10$^{31}\ergs$. The rest of the 2--8 keV emission of
luminosity $L_X = (3\pm 1)\times 10^{34} \ergs$ shows a core-halo
morphology with a core of about 25$\arcsec$ half-width and $\sim
5\arcmin$ extended halo. A weak thermal plasma component of
temperature $\sim$ 0.7 keV was found comprising about 5\% of the
detected diffuse X-ray flux while the dominant emission is due to hard
continuum. The estimated emission measure $\int{}{}n_e n_H$dV is about
$11^{+4}_{-2}\em$ within the 1$\arcmin$  inner circle and
$6^{+2}_{-2}\em$ in the ring between 2$\arcmin$ and 3.5$\arcmin$
providing the plasma number density of $\sim 0.3 \cmc$ in the outer
region. \citet{munoea06} suggested that the hard continuum component
can be fitted either as hot $\gsim$ 3 keV thin plasma with no
prominent He-like Fe line emission requiring a low Fe abundance of
$\sim$ 0.3 of the solar value, or as power-law of photon index $\Gamma
 = 2.7\pm 0.2$ within the inner 1\arcmin and harder index outward with
$\Gamma = 1.7\pm 0.1$ in the 2.5$\arcmin$- 3$\arcmin$ annulus.

To address the dilemma \citet{Westerlund1_XMM_12} analyzed a 48 ks {\sl
XMM-Newton} observation of Westerlund~1 and found He-like Fe 6.7 keV
line emission in the central 2$\arcmin$ region suggesting that the
hard X-rays from the cluster core are mostly thermal in their origin.
They concluded  that  the majority of the hard emission in the inner
2$\arcmin$ annuli likely comes from the thermalized cluster wind
\citep[see also models by][]{oskinova_winds05} with a contribution
from the pre-main sequence stellar population, while it is unlikely
that SNRs are contributing significantly to the diffuse emission of
Westerlund 1 at the current epoch. The Fe line signatures were not
found in the outer extraction annuli by \citet{Westerlund1_XMM_12} and
therefore the question of the thermal or non-thermal origin of  the
hard emission in the outer regions has been left open.

The only clear signature of the presence of an SNR in Galactic compact
clusters of young massive stars is the X-ray pulsar CXO~
J164710.2-455216 (CXOU~J1647-45 afterwards) found in the Westerlund I cluster. The pulsar was
discovered by \citet{NS_WesterlundI_muno06} with {\sl Chandra}
observations. CXOU~J1647-45 has a period of 10.6107(1) s and
the estimated spin down power $\Edot \lsim 3\times10^{33} \ergs$
comparable to its X-ray luminosity, which imply that the pulsar is
likely a magnetar powered by its high magnetic field (expected to be
above 4$\times 10^{14}$ G).  The pulsar is located at about 1\arcmin.6
away  from the core of Westerlund 1 and possibly originates from a
progenitor star of mass above 40 $\msun$. If a neutron star  was born  with a very high magnetic field
(say, $\sim$ 10$^{16}$ G)  with a few ms period
then the strong magnetic dipole radiation losses should slow down its  period to several seconds in a few thousands years  \citep[see, e.g.,][]{mereghetti08}.

Note here that apart from the magnetar CXOU~J1647-45 a number of other soft gamma-ray repeaters, namely
 SGR 0526−66, SGR 1900+14 and SGR 1806−20  are likely associated with
different  massive star clusters \citep[see, e.g.,][for a review]{mereghetti08}. This association may imply the origin of the magnetars from the very massive progenitor stars exploding in young massive star clusters.

Relativistic winds of newly born magnetars  with initial spin rates close to the centrifugal breakup limit were suggested by \citet{blasi_magnetar_ApJ00} and \citet{arons03} to be the sources of ultra high energy cosmic rays. The authors pointed out that a magnetar with a magnetic dipole moment $\mu$ rotating with an angular velocity $\Omega$ has magnetospheric voltage drops across the magnetic field with magnitude $\Delta \Phi \propto \mu \Omega^2/c^2 \approx 3\times 10^{22} (\mu/10^{33}cgs) (\Omega/10^4 s^{-1})^2$ V. \citet{arons03} argued for a hard  power law distribution of light ions  $\propto p^{-1}$ and the spectral steepening to $\propto p^{-2}$ at higher energies, with an upper cutoff energy at 10$^{21}$-10$^{22}$ eV.  In case of the relativistic magnetar magnetosphere the electromagnetic rate of rotational energy loss is close to that of vacuum dipole radiation $\dot{E} = c \Delta \Phi^2$.  The electromagnetic energy emitted by the magnetar may escape the supernova envelope without substantial losses driving a relativistic blast wave in the ISM up to a parsec scale. In order to not overproduce the number of observed interstellar supershells \citet{arons03} suggested that the huge initial spin-down energy of a magnetar is mostly radiated in kilohertz gravitational waves just in several hours after the birth. He also predicted that the gravitational radiation events should correlate with the bursts of UHECRs and that the {\sl  Pierre Auger Observatory} should see the bursts of CR particles with energy above 10$^{20}$ eV once in a few years.
Recently, \citet{lemoine_UHECR_PWN14} proposed that CR protons of energies up to  $4\times10^{20}\,$eV  can be accelerated  at the termination shock of the relativistic pulsar wind for a fiducial pulsar rotation period of one msec and a magnetic field of about 10$^{13}\,$G.

A few magnetars are apparently associated with known supernova remnants.  \citet{vink_kuiper06}  found evidences that the explosion energies of supernova remnants  Kes 73 which is associated with AXP 1E 1841-045, CTB 109 (AXP 1E2259+586) and N49 (SGR 0526-66)  are likely close to the canonical supernova explosion energy of 10$^{51}$ erg, suggesting an initial spin periods of the magnetars to be about 5 ms.

Although no radio-shell or any other standard SNR shell appearance was found
yet, gamma-ray sources were detected in the vicinity of  Westerlund~1
by the {\sl  High Energy Stereoscopic System, H.E.S.S.} in the TeV
band \citep[][]{hessWd1} and by the {\sl Fermi Large Area Telescope}
in the GeV band \citep[][]{ohmea13} with gamma-ray flux $\gsim
10^{-11}\enfl$. \citet[][]{ohmea13} suggested a pulsar wind nebula is contributing to
 the observed gamma-ray emission and concluded that it
provides a possible interpretation of the GeV emission, but can not
explain the emission in the TeV band. On the other hand, they found
that a model with proton acceleration in the cluster's supernova
explosion(s) with subsequent interaction with the surrounding
molecular material may explain both the GeV and TeV emission, but
requires a very high energy input in protons and rather a slow
diffusion.

In \S\ref{CSF} we shall discuss in some detail  another potentially very fast and efficient mechanism
of particle acceleration in colliding shock flows suggested by
\citet{MNRAS_BGO13}.  The colliding shock flow occurs in a compact cluster where a strong expanding supernova
shock is approaching a fast wind of a nearby massive star or collides with the supersonic wind produced by the compact cluster. The colliding shock flows provide the most favourable conditions for the fast CR acceleration by the first order Fermi acceleration. In this scenario PeV regime CRs are efficiently produced on a short time scale of a few hundred years by a supernova remnant interacting with the winds in the compact cluster. Then the accelerated CRs diffuse away on a time scale of thousands years producing both gamma-ray and neutrinos by inelastic interactions of CRs with a matter of ambient clouds.

\begin{figure}
\includegraphics[width=350pt,height=250pt]{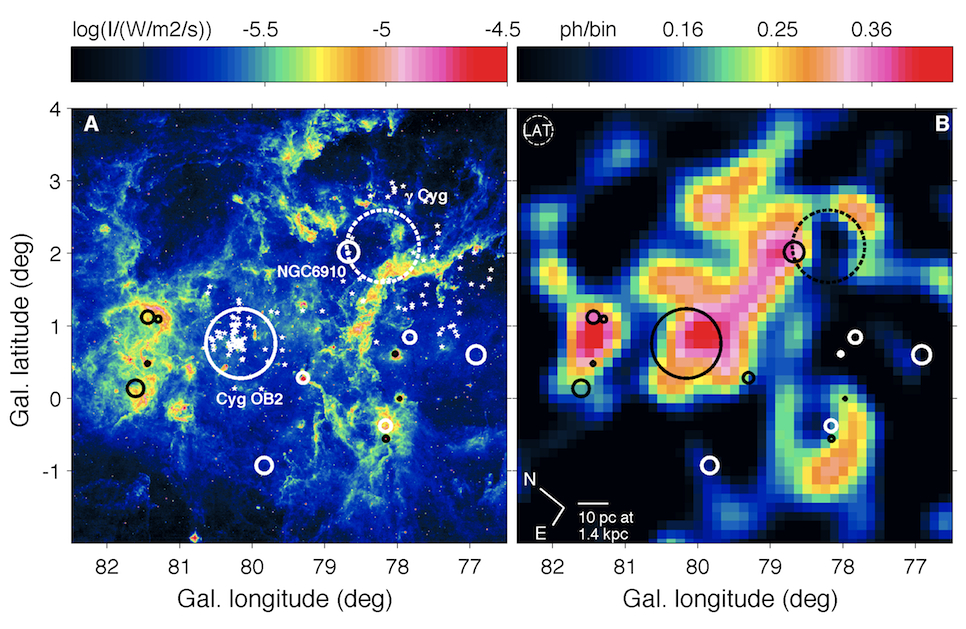}
\caption{{\it Left Panel:}  Mid-infrared 8 $\mu$m intensity map of the Cygnus X region
made with {\sl Midcourse Space Experiment} where rich OB stars association Cyg OB2 (white circle)   and supernova remnant  $\gamma$-Cygni  (white dashed circle) are shown.
{\it Right Panel:}{\sl Fermi-LAT} 10- 100-GeV band photon count map of the Cygnus X region  where a cocoon of freshly accelerated CRs of about 50-parsec-wide  discovered by \citet{fermiSB11}.
Credit: NASA/DOE/Fermi LAT/Grenier Isabelle.}
\label{fig:Fermi_Cygnus}
\end{figure}

\subsection{Non-thermal emission from globular clusters and superbubbles}

Recently \citet{Chandra_Tuc47_ApJ14} re-analyzed {\sl Chandra}
observations of the old Galactic globular cluster 47 Tucanae and
reported a new diffuse X-ray emission feature within the half-mass
radius of the cluster. They fitted the spectrum of the diffuse
emission with a unform thin thermal plasma model of  $T \sim0.2$ keV
and a very hard power-law component of a photon index $\Gamma \sim1.0$
which exponentially falls off outwards from the cluster core. The
non-thermal X-ray luminosity of the region of interest was found to be
$\gsim 10^{32} \ergs$ assuming a 4.5 kpc distance to 47 Tucanae. The
authors proposed two potential scenaria of the origin of the
non-thermal emission: synchrotron/inverse Compton radiation of
particles accelerated by  multiple shocks of colliding winds or
inverse Compton scattering of a pulsar wind. Possible signatures of
non-thermal X-ray components were also reported  from a nearby massive
starforming region ON~2 \citep{oskinova_ON2SFR_nonth10}, and from a
few superbubbles located in the Large Magellanic Cloud \citep[see
e.g.][]{bamba_SB_30Dor_Apj04,maddox_LMC_SB09}.

Detailed study of  30 Dor C superbubble with {\sl XMM-Newton}
was performed recently by \citet{XMM_30Dor14}. Apart from the hard X-ray shell which is highly correlated with the H$_{\alpha}$ and radio shells the  non-thermal X-ray emission was detected from all regions of 30 Dor C as well.  The authors found that the non-thermal X-ray emission may be fitted with a power-law model but an exponential cut-off synchrotron model may fit the spectrum better. The maximal energy of electrons $\epsilon_{max}$ which is required to justify the X-ray synchrotron model was estimated  by \citet{XMM_30Dor14} as $\epsilon_{max} \approx$ 80 TeV $(B/10 \mu G)^{-1/2}$. The synchrotron losses time of the electrons is below 1.5 kyrs and therefore efficient particle acceleration is required.  \citet{XMM_30Dor14} argued that 30 Dor C is currently undergoing a phase of high energy particle acceleration by massive star winds and perhaps supernova remnants expanding in the hot superbubble. The maximal energy of the electrons might be somewhat lower than that derived by \citet{XMM_30Dor14} if one account for the effect of turbulent magnetic fields in the superbubble on the spectra of the synchrotron radiation .

The spectrum of the giant starforming region 30~Doradus in the LMC in
the energy range from 100 MeV to 100 GeV obtained with the  {\sl Fermi
LAT} is shown  in  Fig.~\ref{fig:30Dor_Fermi} (the
blue data points).  The spectrum of the source shown in the right
panel  is consistent with a photon index $\Gamma \approx 2$ and the
estimated  gamma-ray luminosity is $\gsim 4\times 10^{36} \ergs$
\citep[see, e.g.,][]{LMC_Fermi_AA10}. The luminosity is well below the
gamma-ray luminosities of 10$^{39}$ to 10$^{41}\,\ergs$, which are
implied from recent observations of the powerful starburst galaxies M82, NGC 253,
NGC 1068, NGC 4945 \citep[see,
e.g.,][]{Fermi_M82_NGC253_ApJ10,Fermi_SF_ApJ12}
(they are indicated in Fig.~\ref{fig:SFG_Fermi_TeV}), but it is
substantially larger than the 1--100 GeV luminosity of the Cygnus
cocoon discovered by  \citep{fermiSB11} as discussed below.

The {\sl Fermi LAT} telescope discovered a cocoon of about 50 parsec width located in between
Cygnus~OB2 and $\gamma$-Cygni SNR with a peak of the emission close to
the massive-star cluster NGC 6910 \citep[][]{fermiSB11}. The cocoon
in Figure~\ref{fig:Fermi_Cygnus} shows signs of enhanced fluxes of low energy cosmic rays likely
powered by multiple shocks of fast stellar winds and SNRs  located in
the Cyg OB2 region, which overlaps with a TeV source MGRO J2031+41.
Pulsar wind nebulae expected in the region with multiple SNRs could
also contribute to these fluxes, but  \citet{fermiSB11} pointed
out that the young pulsars J2021+4026  and J2032+4127 are unlikely to
explain the Cygnus cocoon emission.  The gamma-ray
luminosity of the cocoon in the 1--100 GeV range is about $(9\pm2)
\times 10^{34} \ergs$ assuming the 1.4 kpc distance to the Cygnus OB2
region. Recently  {\sl ARGO-YBJ} reported an identification of the extended
TeV source  ARGO J2031+4157 ( MGRO J2031+41)  with the Fermi Cygnus cocoon.
They fitted 0.2-10 TeV  spectrum of ARGO J2031+4157 with a photon index $\Gamma =  2.6 \pm 0.3$  \citet{Cygnus_cocoon_ARGO-YBJ14} while {\sl Fermi LAT} spectrum below 100 GeV seems to be
better fitted with a flatter power-law photon index $\sim$ 2.2. All of these systems of
different spatial scales and injected kinetic power are characterized
by  spectra of photon indexes 2--2.5  in GeV--TeV regimes studied with {\sl Fermi LAT} and ground-based TeV
telescopes.

Gamma-ray observations in the TeV regime with ground based atmospheric
Cherenkov telescopes {\sl H.E.S.S.},  {\sl MAGIC},  {\sl VERITAS}, and
 {\sl Milagro}, a water Cherenkov detector reaching a high
sensitivity, revealed a number of sources in the Cygnus region
\citep[see, e.g.,][]{aharCygOB_02,aharon_Survey06,MAGIC_Cyg08,VERITAS_Cygnus13,Westerlund2_HESS11}
and of young massive star clusters \citep[see, e.g.,][]{hessWd1,
Westerlund2_HESS11}. Some of the sources are likely extended. While
the exact identification of the extended sources is still to be made
with the multiwavelength observations, the most plausible associations
are related to either pulsar wind nebulae or multiple shock systems
within OB star clusters and SNRs  \citep[see, e.g.,][and references
therein]{buttea03,Torres_Reimer13}.

\section{Hydrodynamical models of massive star winds and superbubbles}\label{SBsHyd}

To interpret adequately the multi-wavelength observational data on the
starforming and starburst regions one has to rely on models. The
evolution of an SF system is an integral part of the ISM ecology and
it starts from molecular cloud formation and ends with the destruction
of the cloud by radiation and winds of massive stars, supernova
explosions, dispersion of open clusters of massive stars or massive
globular cluster formation. Above we have briefly discussed some
aspects of the first stages of the SF process and a more detailed
discussion can be found in
\citet{bromm_larson_first_starsARAA04}, \citet{mckee_ostriker_e_ARAA07}, \citet{massive_star_form_2007ARA&A}, \citet{elmegreen11a}, \citet{kennicuttARAA12,SF_AARv14}, \citet{Krumholz_SFR_PR14}, \citet{2014arXiv1402.6196M}, and  \citet{massive_star_tan_ProtSP14}.

\begin{figure}
\includegraphics[width=350pt]{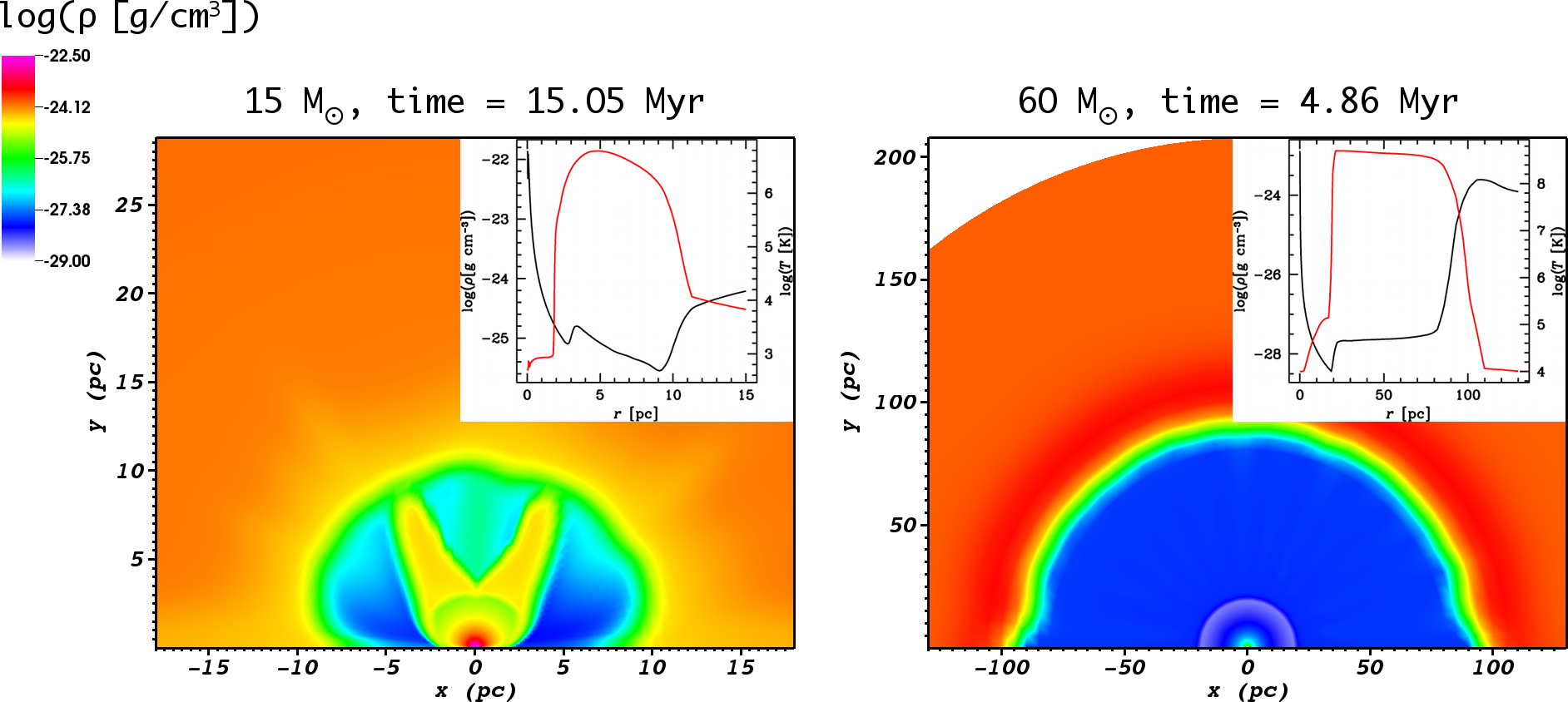}
\caption{Typical pre-SN aspect of the circumstellar matter around a $15\,\Msun$ star
\textit{(left panel)} ending its stellar life as a red supergiant (and exploding
as a type IIP SN), and around a $60\,\Msun$ star \textit{(right
panel)}, ending as a Wolf-Rayet star (exploding as a type Ibc SN). In the
small windows the mean density (black solid line)  and temperature
(red solid line)  are shown as a function of the radius. Adopted from
\citet{Georgyea14}.} \label{wind_Georgy}
\end{figure}

Basic physics of the mechanical and thermal equilibrium of rotating massive stars  and their evolution has been thoroughly reviewed by \citet{Maeder_Meynet_RMP12} with implications to population
synthesis models \citep[see, e.g.,][]{Georgyea14}. Radiative forces of high luminosity young massive stars are accelerating  near-surface layers initiating  fast winds and stellar mass loss. Simulations of
the line-driven stellar winds require a careful study of the instabilities at different scales, clumping and porosity effects \citet{abbott82_rad_line_driv_winds, owockiea_rad_driv_wind88, kudritzki_puls_ARAA00,puls_AARv08,owocki_winds14} to model the dependence of mass loss and  wind terminal velocity on the metallicity and stellar rotation and to establish the maximal mass limit and pursue evolution of young massive stars.

The mass loss rates of young massive stars are typically within the range  $10^{-7} < \Mdot <  10^{-4}\, \sfr$ (and may be as high as  5 $\times 10^{-4} \sfr$ for a relatively short luminous blue variable phase)  with the terminal velocities reaching 3,000 $\kms$. For example, the total kinetic energy input from a wind of a 60 $\msun$ star is about 8 $\times 10^{50}$ ergs for a mean terminal velocity of 2,000 $\kms$ before the supernova explosion. The models of stellar evolution by \citet[][]{chiosi_maeder_ARAA86, maeder_meynet_AA88, lamers_cassinelli99} predict that such a star would eject into the interstellar medium about 29  $\msun$  of hydrogen, $\sim$ 8  $\msun$  of He, and  $\sim$  1  $\msun$  of C and O (almost 38 $\msun$  of gas in total). A very useful recipe to estimate the mass-loss rate as a function of the stellar mass and luminosity, effective temperature and terminal velocity of the wind for different metal abundances can be found in \citet{vink_JS_mass_loss00,vink_JS_mass_loss01}. Magnetic fields in OB-stars  are thought to be at the level of $\sim$ 100 G. They play a role in the structure and evolution of young massive stars maintaining a rotation law to be close to uniformity \citep[see, e.g.,][]{magn_field_Maederea09}.  The magnetic coupling is essential to regulate the interconnection between the differential rotation (which it tends to suppress) and the meridional circulation. The magnetic braking effect reduces stellar rotation if the star's mass loss is strong.

The origin of magnetic fields of different scales in massive stars is now a subject of discussions. The dynamo-type models are being considered along with a stellar merging process where the binary evolution may result in a rapid mass transfer phase accompanied by magnetic field amplification.
This may account for magnetic fields of about 10\% of O-type stars which are thought to be produced via merging of two main sequence stars \citep[see e.g.][]{langer_magn_fields13}.

\begin{figure}
\includegraphics[width=300pt]{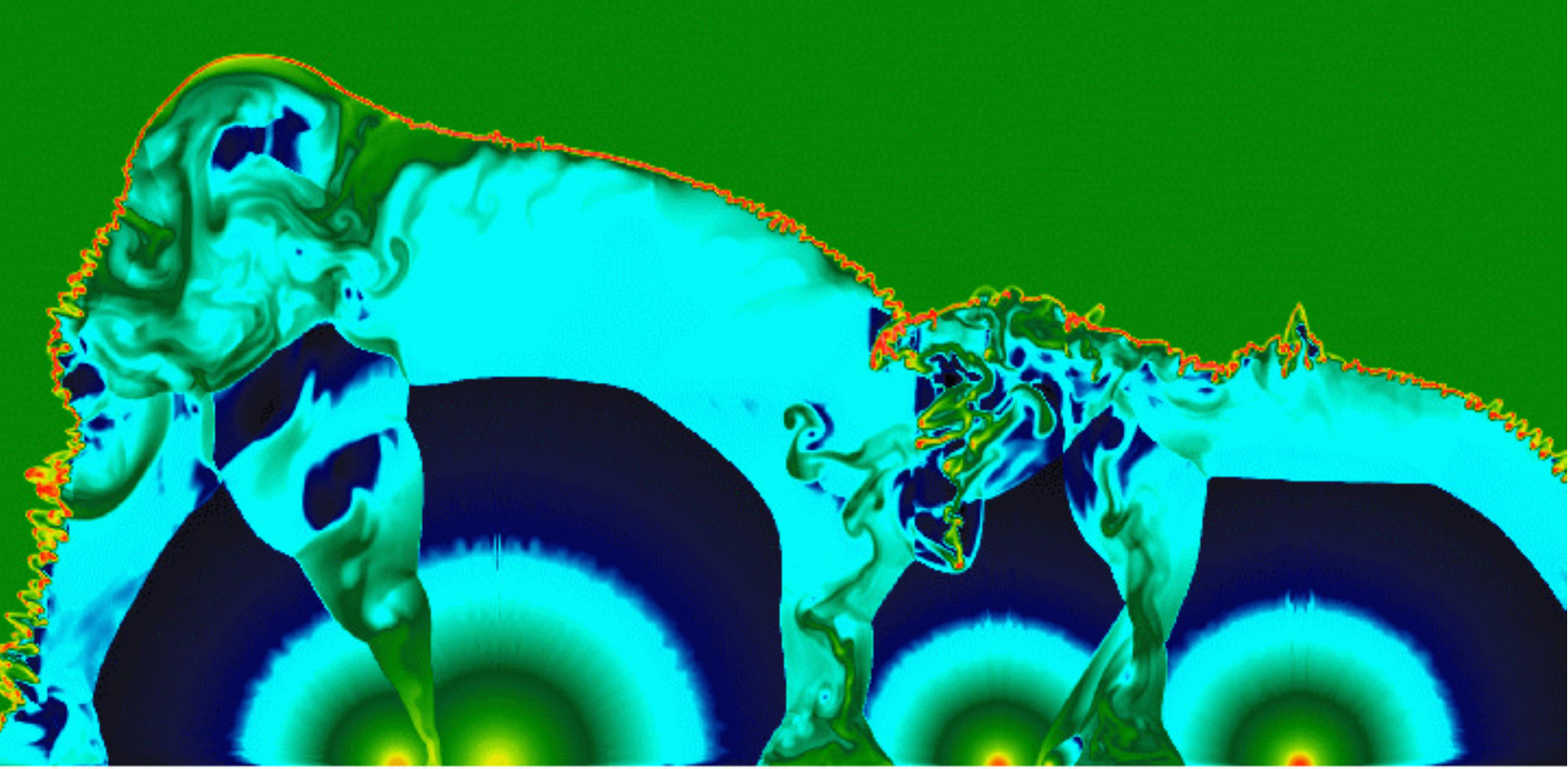}
\caption{Hydrodynamical simulation of an axisymmetric mini-star cluster of 5 stars (from \citet{Georgyea14}). Thin shell instabilities develop in regions of interactions of different winds and of wind-interstellar medium. Gas densities are shown in color: 10$^{3} \cmc$  (white),10 $\cmc$  (red), 1 $\cmc$  (green), 10$^{-2} \cmc$ (blue).  } \label{SB_sketch}
\end{figure}

Massive early type star winds inject energy and momentum into the surrounding medium to blow bubbles of different  scale sizes surrounded by shells  \citep[see, e.g.,][]{castorea75, weaverea77, lamers_cassinelli99, georgyea_13}. Below we will briefly review some recent models of bubbles formed by individual massive early type stars as well as superbubbles produced by collective actions of clustered massive stars and supernovae on the time scale of $\gsim$ 10 Myrs and compare them with available observational data depicted by \citet{chu_bubbles_SBs_review08} who discuss OB associations, hot interiors of superbubbles, swept-up dense shells,  and interfaces between dense shells
\citep[see also][]{chandra_IR_feigelsonApJS13}.

In Figure~\ref{wind_Georgy} wind structure is illustrated along with the recent results of two-dimensional axisymmetric hydrodynamical simulations of the circumstellar medium around 15 $\msun$  and 60 $\msun$ stars at rest taken at the pre-supernova stages perfomed by \citet{georgyea_13}. The  simulations were made for a grid of massive stellar models from 15 to 120 $\msun$ the stellar evolution was followed from the zero-age main-sequence stage to the stage of supernova explosion. Gas velocity, density and temperature profiles are shown in the insets (for details see \citet{georgyea_13}). Fast wind of velocity  of a few thousands $\kms$  is confined in a region of a few pc size and terminated at a strong shock where the tenuous gas is heated up to temperatures  10$^6$--10$^8$ K and the hot bubble may extend for tens of parsecs. The bubble in the simulation expands slower than it is predicted by the self-similar model and its size depends on the metallicity of the central star. The low-metallicity stars have relatively smaller bubbles because their winds are weaker. The effect of the stellar motion on the structure of the circumstellar medium may be important
and was studied by \citet{wind_in_motion14} who have calculated a grid of models of bow shocks around main sequence and evolved massive runaway stars.

Since the young massive stars are highly clumped in the region where they originate, and the mean distances between the stars are typically smaller than the sizes of individual star bubbles, the structure of the circumstellar medium in such a region would be more complex. \citet{velazquez_SNR_wind03} performed a numerical hydrodynamic simulation of the interaction of supernova ejecta with a stellar wind. They modeled the X-ray emission from a supernova remnant --- stellar wind collision region and found shell-like structures of enhanced X-ray emission. It is instructive to note that the simulated system at some evolutional stage (of about some thousands of years duration) may become a very efficient accelerator of energetic particles. It may transfer about a half of the kinetic ram pressure into the energetic non-thermal particles as demonstrated within the one-dimensional non-linear kinetic model of \citet{MNRAS_BGO13}.

A 2D hydrodynamical model of the temporal evolution of an aspherical circumstellar bubble created by two massive stars  of 25 $\msun$ and 40 $\msun$  separated by 16 pc in a cold 20 $\cmc$ medium, which included the stage of supernova explosion was presented by \citet{marleea12}.  In Figure~\ref{SB_sketch} adopted from \citet{georgyea_13} we illustrate density distribution at some intermediate stage of a simulated 2D evolution of a cluster of 5 massive stars.

Evolution and X-ray spectra of  three massive star system  of  25,  32, and 60 $\msun$  in a background matrix of 10 $\cmc$ density was studied by \citet{krause13_SB1,krause14} in 3D hydrodynamics with an account for optically thin radiative cooling and photo-electric gas heating. They found that the energy injected by
supernovae is entirely dissipated in a superbubble on the timescale of about 1 Myr, and that the  deposition from the stars to the superbubble did not vary substantially if the massive stars were localized within a distance of $\lsim$ 30 pc. The simulations showed the outer dense shell thickness to be  $\sim$ 0.1 of the superbubble size in accordance with observations. The X-ray luminosity of a superbubble increases by a factor $\lsim$ 100 for a period of about 0.1 Myr after a supernova explosion reaching a peak luminosity $\sim$ 10$^{36}\ergs$. \citet{rogers_pittard_SNe_inYMCs13, rogers_pittard_SNe_inYMCs14} simulated distraction of a giant molecular cloud clump of about 3240 $\msun$ mass by three O-stars localised within a 4 pc radius.

The simulations demonstrated a highly structured environment into which the supernova energy is released after 4.4 Myrs of the system evolution, allowing most of the supernova energy to reach wider environments.The molecular component is  completely destroyed and dispersed in 6 Myr. The authors found that the X-ray luminosity of the system is about 10$^{34}\ergs$ at the early stages of the evolution and during short periods of Wolf-Rayet stages. However, it goes down to about 1.7$\times 10^{32} \ergs$ between 1--4 Myrs and it can increase to $\sim 10^{37}\ergs$ for relatively short time  periods $\gsim$ 4,600 yrs after each supernova explosion. Because of the leakage of hot gas material through gaps in the outer shell, the X-ray luminosities obtained in 3D simulations by \citet{rogers_pittard_SNe_inYMCs14} are
generally considerably fainter than those predicted within spherically-symmetric bubble models.

 \citet{maclow_mccray88} presented a general view on the ISM in the vicinities of starforming regions where correlated supernova explosions in OB associations create a superbubble: a large, thin, shell of cold gas surrounding a hot pressurized interior. The dynamics of a superbubble was approximated as that of a source of continuous mechanical luminosity assuming that supernova blast waves become subsonic before they reach the walls of the shell. A thin shell (the Kompaneets' approximation) is widely used \citep[see, e.g.,][]{BK_silich95} to describe supershell propagation in inhomogeneous interstellar medium. Superbubbles may be blown out of the H I layer into the halo. The evolution of superbubbles in an exponential ISM density distribution is a subject of recent analytic and numerical studies  \citep[see, e.g.,][]{b_breitschwerdt13, CR_driven_gal_wind13}. In this context \citet[][]{db_gwinds12}  suggested a model which can explain the power-law distribution of CRs between the ``knee'' and the ``ankle''. The impact of supernovae in superbubbles on the galactic ecosystem was
discussed in various aspects by \citet{SNe_feedback_SFR_Hennebelle14} and by \citet{wang14}. One should have in mind the existence of a feedback effect of supernova and superbubbles on the global galactic ecology. This is the effect of large scale galactic magnetic field amplification by the cosmic-ray-driven dynamo on the timescale of about 150 Myr simulated by \citet{hanasz09, CR_driven_dynamo_AA14}. \citet{MF_cluster_formation_MNRAS09} demonstrated that even a very low magnetic field is able to significantly influence the star formation process. Global properties of
superbubbles can be affected seriously by the local effects of magnetic field, thermal conduction, turbulent mixing as well as by energy exchange with accelerated particles which are connected with the thermal plasma  motions via fluctuating magnetic fields produced by CR-driven instabilities \citep[see,
e.g.,][]{bell04, Zweibel_Everett10, ber12, schureea12, blasiAARv13} and may comprise a sizeable fraction of the free energy of a superbubble \citep[][]{b01, bb08}.

 \section{Observational appearance of massive star winds and superbubbles: the non-thermal emission}

The structure of the circumstellar medium modified by a strong wind of a massive star and a presence of molecular clouds seriously influence both the thermal and the non-thermal appearance of a supernova remnant \citep[see e.g.][]{chevalier99, chevalier05, chevalier14, bceu00, ellisonea12}. The structure of magnetic fields in the winds of massive stars was discussed by, e.g., \citet[][]{walderea12}. The magnitude of magnetic field in the close vicinity of the progenitor star may be high enough to allow acceleration of particles in very young supernova remnants to energies well above the PeV regime \citep[e.g.,][]{vb88, Ptu10}. On the other hand, magnetic field in the outer regions of the wind of a slowly rotating massive star may drop below a $\mu$G level. The low magnetic field  allows the inverse Compton interpretation of the observed GeV-TeV  gamma-ray photons from synchrotron X-rays dominated supernova shell RX~J1713.7-3946 in the frame of the diffusive shock acceleration model with strong magnetic field amplification. In this model the low magnetic field in the outer region of pre-supernova wind is amplified by CR driven instabilities by a factor of $\sim$ 50 (say, from 0.4 $\mu$G to 20 $\mu$G) \citep[][]{ellisonea12}.  The gamma-ray image of this SNR morphologically resembles the synchrotron X-ray map \citep{RXJ1713_HESS_AA07}. The models, which explain the origin of the gamma-ray photons via the inverse Compton scattering of background photons by the synchrotron X-ray emitting electrons  typically require magnetic field in the emission region to be $\sim$ 20 $\mu$G \citep[see, e.g.,][for a discussion]{vink12}.  This is difficult to reconcile with an efficient particle acceleration at supernova shock, which results in a high magnetic field amplification factor (above 10) if the shock is propagating into the standard ISM with magnetic field of a few $\mu$G observed under the typical ISM conditions \citep[][]{beck12, heiles_haverkorn12}, unless the SNR is expanding into the wind of a pre-supernova.

\subsection{Colliding wind binaries}

The wind collision in early-type binary systems is the next level of complexity.  The  wind structure, position and  shape of the contact discontinuity between the stars,  double-shock structures,  as well as the post-shock flows were examined by \citet{stevens92} and later by \citet{parkin_pittard_3D_coll_winds08}, who correspondingly used 2D and 3D hydrodynamic codes with an account for radiative cooling and some instabilities. The models were aimed to describe the symbiotic systems and gamma-ray binaries, and also binary systems with O-type and Wolf-Rayet stars including the enigmatic object in $\eta$-Carinae. The colliding wind binaries were long known to be sources of non-thermal radio emission \citet{eu93, dougherty06, pittard_dougherty06, deBecker_review07} and the catalog of colliding wind binaries with non-thermal particle acceleration compiled by \citet{deBecker_catalogue_CWB13} contains 43 objects.

\citet{dougherty06} presented milliarcsecond resolution {\sl Very Long Baseline Array} observations of  the colliding-wind binary WR~140 (HD~193793) at different orbital phases. WR~140 is a massive star binary with estimated masses of 20$\pm$ 4 $\msun$ for the W-R star and 54$\pm$10 $\msun$ for the O star and a period of 7.9 days. The system has a highly elliptical orbit where the stellar separation varies between 2~AU at periastron and about 30~AU at apastron. The mass loss rate determined from  the thermal free-free emission of WR wind is $\gsim 2\times 10^{-5}\sfr$ being a subject of uncertainty due to likely clumping of the wind of a terminal velocity $\sim$ 3,000 $\kms$. The estimated total kinetic luminosity of  WR~140 is  $\sim 6\times 10^{37} \ergs$ while the radio  luminosity is $\sim 2.6\times 10^{30} \ergs$ \citep[see][]{deBecker_catalogue_CWB13}. The total radio flux measured with VLBA does not change from one orbit to the next, but shows strong variations with the orbital motion exhibiting both optically thin and optically thick synchrotron radio emission  \citep{dougherty06}. \citet{pittard_dougherty06} estimated the kinetic power in the wind -- wind collision to be $\sim 5 10^{36} \ergs$ of which -- they suggested -- about 0.5\% is being transferred to the broad band non-thermal emission observed from radio to gamma-rays, which is due to energetic particles accelerated by shocks \citep[c.f.][]{eu93, bednarek_DSA_CWB05} or magnetic field reconnection. By now {\sl Fermi LAT} observations provided an upper limit of $\sim 9.6\times 10^{-9}\fl$ (for one energy bin between 95.6 MeV and 44.9 GeV) for the gamma-ray emission from WR~140 \citep{Fermi_limit_CWB13}. The same authors established upper limits on gamma-ray emission from a number of other colliding wind binaries.

Gamma-ray emission and hard X-ray non-thermal emission was detected from $\eta$-Carinae  \citep{Tavani_eta_Car_ApJ09, Fermi_bright_source_list09, eta_Car_Farnier_AA11, dubus_gamma_ray_binariesAARv13} which is considered as a likely case of colliding wind binary. No clear evidence for a non-thermal synchrotron component in the observed radio emission of $\eta$-Carinae has been reported yet, possibly due to a high free-free absorption \citep[see, e.g.,][and references therein]{deBecker_catalogue_CWB13}. An analysis of broad-band X-ray, optical and near-infrared observations carried out within about 60~yrs \citep{eta_Car_periodMNRAS08} revealed the binarity of the system with the period of 2022.7 $\pm$ 1.3 days (about 5.54 yrs). The total mass of the system is likely $\gsim$ 110 $\msun$. The primary star $\eta$-Carinae A is  likely a luminous blue variable star which is characterized by a great mass loss rate estimated as  $2.5\times$ 10$^{-4}$ -- 10$^{-3}\sfr$ and the wind terminal velocity $\lsim$ 600 $\kms$ with giant eruption events in the past that may have ejected some material at velocities about 3,000 $\kms$. The probable companion star named $\eta$-Carinae B has not been directly observed, but the analysis of X-ray observations hinted on the presence of its wind of high terminal velocity $\gsim$ 3,000 $\kms$ and the mass loss rate  $\sim 10^{-5}\sfr$ \citep{pittard_corcoran_eta_Car02, eta_Car_Groh_AA10}. The system is located at the distance of 2.3 kpc and has a complex geometry of winds and shells \citep[see, e.g.,][]{Madura_eta_Carinae14}.

Analyzing  {\sl Fermi LAT} data on the source FGL~J1045.0-5942 associated with $\eta$-Carinae together with {\sl INTEGRAL IBIS} observations \citet{eta_Car_Farnier_AA11} suggested a two-component gamma-ray spectrum of the source with a hard component of photon index $\Gamma$ = 1.69 $\pm$ 0.12 with exponential cut-off at $\sim$ 1.8 $\pm$ 0.5 GeV present in the energy range 0.2 -- 8 GeV. The authors suggested the leptonic origin of the component from the inverse Compton upscattering of the background photons. The second component is a power-law of index $\Gamma$= 1.85 $\pm$ 0.25 in the high-energy regime 10 -- 100 GeV with the normalization flux lower than that of the hard one; it was attributed to the hadronic component due to pion decays. The total gamma-ray  luminosity of FGL~J1045.0-5942 is about 10$^{34} \ergs$.

\section{Particle acceleration and non-thermal radiation in starburst complexes}

Non-thermal emission detected in the broad band spectra of colliding wind binaries, supernova remnants,  young massive star clusters, superbubbles and starburst galaxies  are the signatures of violent energy release in the systems and of particle acceleration. Intense plasma motions powered by gravitation, anisotropic radiation flows, and magnetic fields may transfer a part of the available free energy to a population of highly non-thermal particles. The efficiency of the energy conversion  depends on the particular acceleration mechanism and may reach tens of percent. Two kinds of particle acceleration processes are evidently at work in different astrophysical objects: those related to magnetic field reconnections and to magnetohydrodynamic shocks and turbulence.

To release the free energy of highly magnetized plasma systems with globally frozen-in magnetic fields effective dissipative processes have to work in a regime with infrequent Coulomb collisions as well. Electric fields associated with the magnetic field reconnection/dissipation regions which allow to change the field topology accelerate charged particles and heat plasma.

The second type of process is most likely operating in starburst complexes given that the radiative pressure driven winds of massive stars are not highly magnetized and that fast MHD shocks are the most common phenomena in supersonic flows of colliding winds and supernovae. Before a discussion
on particle acceleration by shocks and turbulence we will briefly summarize some facts concerning particle acceleration during magnetic reconnection.

 \subsection{Particle acceleration due to magnetic field reconnection}

Efficient particle acceleration may occur in the systems where the magnetic field configuration storages free energy, for example in stellar atmospheres, magneto-tails of fast moving stars, highly magnetized winds or jets. Rapid reconnection of magnetic fields of different topologies may happen in spontaneous current sheets or be due to an external driving force \citep[see, e.g.,][]{zweibel_yamada_reconn_ARAA09}. Mechanisms of direct acceleration of charged particles by the reconnecting magnetic field were studied in detail both analytically and numerically \citep[see, e.g.,][]{Zelenyi90, Hoshino05,birn_priest07,Drake06, Pritchett06, lazar_opher_ACR_recon09,Artemyev13, birnea12}.

An early analytical study of test particle orbits around X-line field topology with and without a guide field (where the electrons remain magnetized) was performed by \citet{bs76} and later advanced with Particle-in-Cell (PIC) technique \citep[see, e.g.,][]{oka_reconn_ApJ10, bessho_reconn_ApJ12, ss_PICreconn_12}. \citet{bs76} came up with an exponential asymptotic shape of the energetic electron spectra in both relativistic and non-relativistic regime. In the relativistic regime $N(\gamma) \propto \exp{(-(\gamma/\gamma_m)^{1/2})}$ where the electron Lorentz factor $\gamma \gg$1 and the maximal Lorentz factor $\gamma_m$ can be roughly estimated as $\gamma_m \sim  \delta_{\rm s} \times \epsilon^2 $. Here $\epsilon$ is the ratio of the electric field inside the X-line to the amplitude of magnetic field at the boundary of the current sheet system, whose geometry is characterized by the dimensionless current sheet stretching $\delta_{\rm s}$. Under typical current sheet conditions $\epsilon \sim 10^{-3}-10^{0}$, while $\delta_{\rm s} \sim 10^4-10^6$ for the electrons and $\delta_{\rm s} \sim 10^3-10^4$ for the ions. Recent two-dimensional PIC simulations of magnetic reconnection and particle acceleration in relativistic Harris sheets in low-density electron-positron plasmas with no guide field performed by \citet{bessho_reconn_ApJ12} revealed particle spectrum
$N(\gamma) \propto  \gamma^{-1/4}\, \exp{(-(\gamma/\gamma_m)^{1/2})}$.

Due to spatial and temporal localization of a magnetic reconnection region the efficiency of charged particle acceleration is substantially limited. In particular, formation of exponential tails of accelerated particle distributions results from the instability of charged particle motion in the vicinity of magnetic field null points. The amounts of accelerated particles and their energies can be significantly increased in case the of multiple reconnection when one considers formation of several magnetic islands interacting with each other \citep{Hoshino12}. Transient magnetic reconnection in hot plasma is responsible for generation of strong plasma jets, which accelerate electrons via adiabatic mechanisms \citep{Birn04, AshourAbdalla11}. Hall electric fields play an importnat role in electron trapping and acceleration in the vicinity of a reconnection site \citep{Hoshino05, Zharkova11}.

Additional effects responsible for charged particle acceleration within a reconnection region are due to resonant and nonresonant charged particle interaction with electromagnetic turbulence generated by plasma flows originated from the reconnection site \citep[e.g.,][]{Grigis_Benz06, pb08, bf09}. Strictly speaking, only models including turbulence or multiple reconnections pretend to reproduce the power law energy distribution of accelerated particles.  In these models piece-wise power law distributions of accelerated particles may have hard spectral indexes due to fast acceleration rates and extended acceleration regions. The essential part of such models corresponds to stochastic charged particle interaction with magnetic field structures representing magnetic islands, current filaments and strong waves.

It was recently established that magnetic field reconnection effects are likely responsible (under some specific conditions) for dissipation of fully developed magnetic turbulence.  In the scenario considered by \citet{turb_diss_reconn_Servidio11} spontaneous magnetic reconnection is locally driven by the
dynamical forces under boundary conditions provided by the turbulence.

With kinetic simulations
that span a wide dynamical range from the macroscopic fluid scales down to the electron scales \citet{karimabadi13} demonstrated  that a turbulent cascade results in formation of coherent structures of multiple current sheets that steepen to the electron scales. This results in strong localized heating of the plasma due to magnetic reconnection and the dominant heating mechanism is related to parallel electric fields within the current sheets leading to anisotropic distributions of electrons and ions. Plasma heating by current sheets was found to be locally several orders of magnitude more efficient than wave damping. This mechanism may be very important to understand plasma heating due to dissipation of strong collisionless magnetic turbulence in fast winds of massive stars, colliding binary systems, supernova shells and superbubbles. By now dissipation of collisionless turbulence has been treated within MHD models at best by introducing some ad hoc phenomenological macroscopic viscosities and an important issue of the electron to ion heating rate ratio shall be addressed with the new approaches to come.

\subsection{Particle acceleration by collisionless shocks}

Shock waves are thought to be the main gas heating and particle acceleration agents in astrophysical flows. We shall discuss below some of the properties of collisionless shocks with an accent on the production of non-thermal components in the form of energetic particles, fluctuating magnetic fields and radiation, which are rather common because of the very slow rate of the Coulomb relaxation process in this type of flows. The large number of degrees of freedom in the astrophysical collisonless shocks determines their multi-scale nature.

It was established more than fifty years ago both observationally and theoretically that collisionless shocks do exist in the mildly magnetized plasmas \citep[see, e.g.,][]{Sagdeev66,tk71}. Nevertheless,
a non-trivial question still remains -- whether the collisionless shocks may occur in a hot unmagnetized plasma, as well as in the highly magnetized plasmas. Two-dimensional PIC simulations of the structure of  nonrelativistic collisionless shocks in unmagnetized electron-ion plasmas by \citet{kt08,kt10} revealed that the energy density of the magnetic field generated by a Weibel-type instability within the shock transition region typically reaches 1\% - 2\% of the upstream bulk kinetic energy density. The width of the shock transition region was found to be about 100 ion inertial lengths, independent of the shock velocity. The hybrid plasma simulations with kinetic treatment of ions and fluid electron description \citep[see, e.g.,][]{wq88,winskeea90,giacaloneea97,treumann09,gs12,bs13} allow to study domains of some thousands gyroradii of upstream protons around non-relativistic shocks. Two-dimensional hybrid simulations of a quasi-parallel shock with the Alfven Mach number $M_a$=6 by \citet{gs12} revealed a power law distribution of energetic ions of index about $2 \pm 0.2$ in the shock downstream. The modelled energetic particle population contained about 15\% of the upstream flow energy. Recent 3D hybrid simulations of non-relativistic collisionless strong shocks with $M_a$=6 in the domain of 2000$\times$200$\times$200 ion inertial lengths performed by \citet[][]{CS2013} showed magnetic field amplification (MFA) in the shock upstream and particle acceleration effects.

\begin{figure}
\includegraphics[width=180pt,height=180pt]{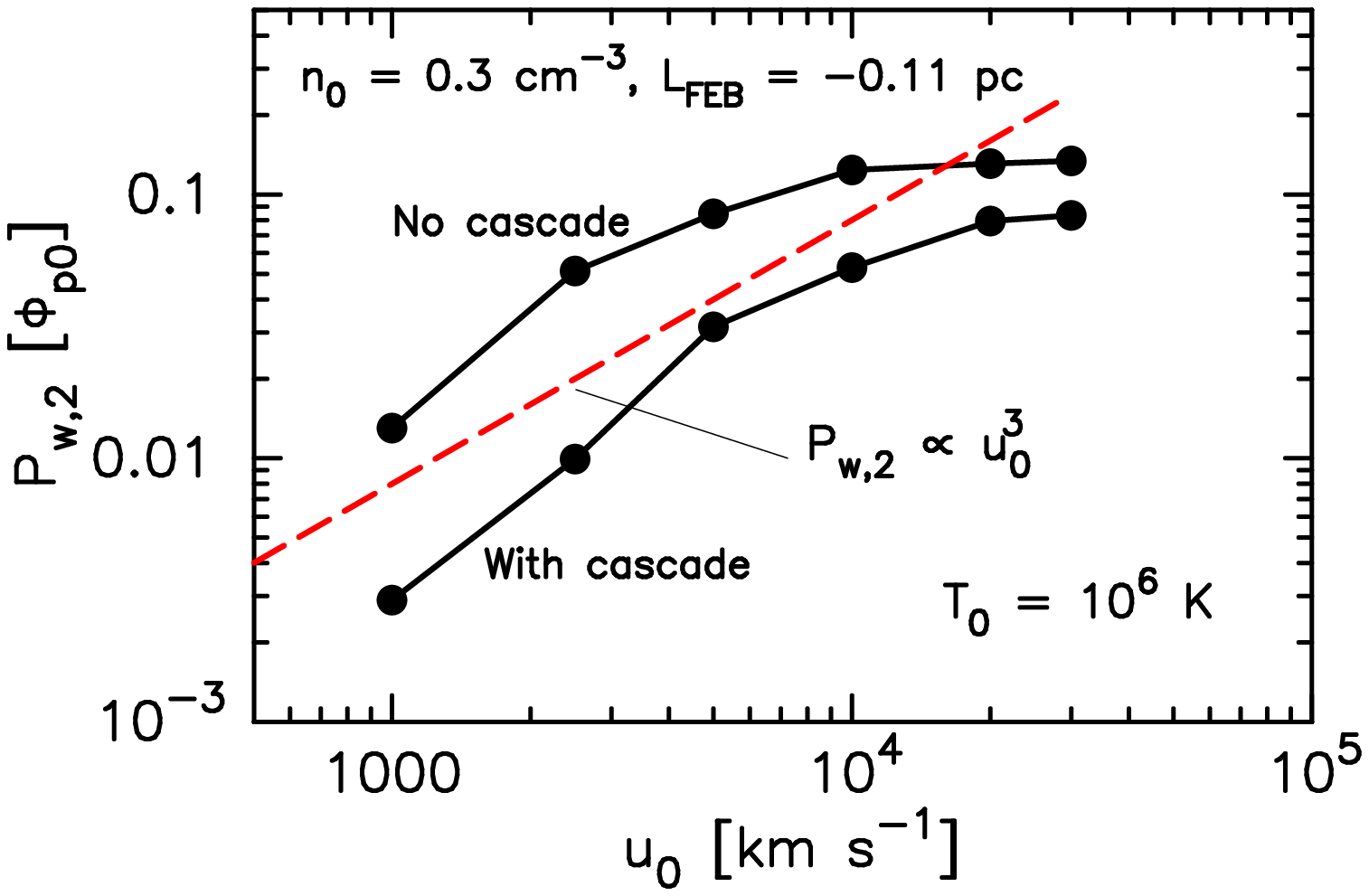}
\includegraphics[width=180pt,height=180pt]{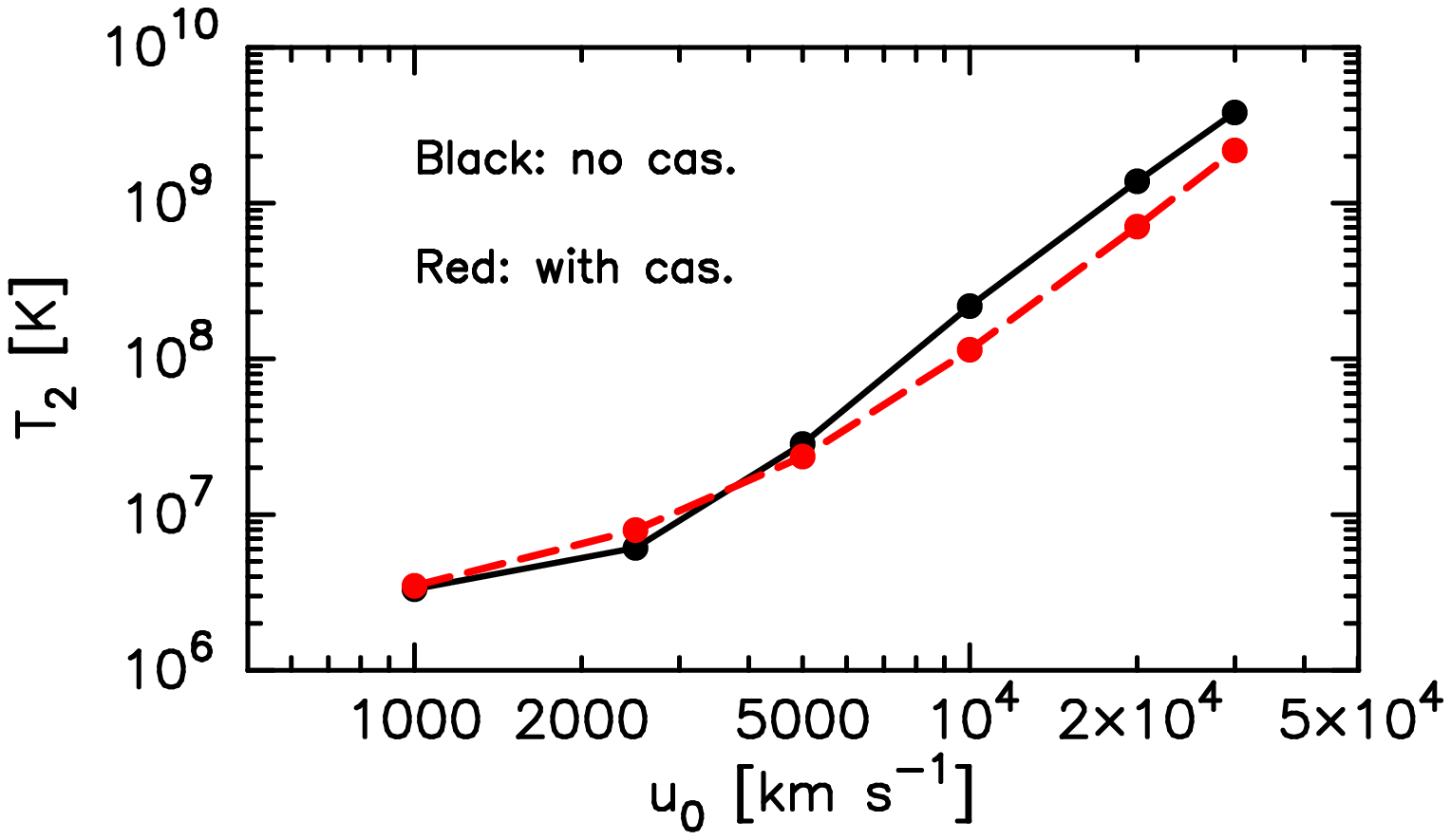}
\caption{{\it Left Panel:} Magnetic field amplification factor in the shock downstream ${\rm P}_{\rm w,2} = B_{\rm t}^2/8\pi \rho_0 u_0^2$ as a function of shock velocity $u_0$  simulated in non-linear Monte Carlo diffusive shock acceleration model by \citet{beov14} with  turbulent cascade and without turbulent cascade. Shocks propagate in plasma with initial number density $n$=0.3$ \cmc$ and T= 10$^6$K.     {\it Right Panel:} The downstream proton temperatures as a function of shock velocity for the same models as that in the left panel. The  departures from the single fluid scaling $T_2 \propto u_0^2$ evident at low velocities are due high gas compression ratio because of kinetic energy escape with CRs in the DSA models.}
\label{fig:DSA:MFA_T2}
\end{figure}

\begin{figure}
\includegraphics[width=180pt,height=180pt]{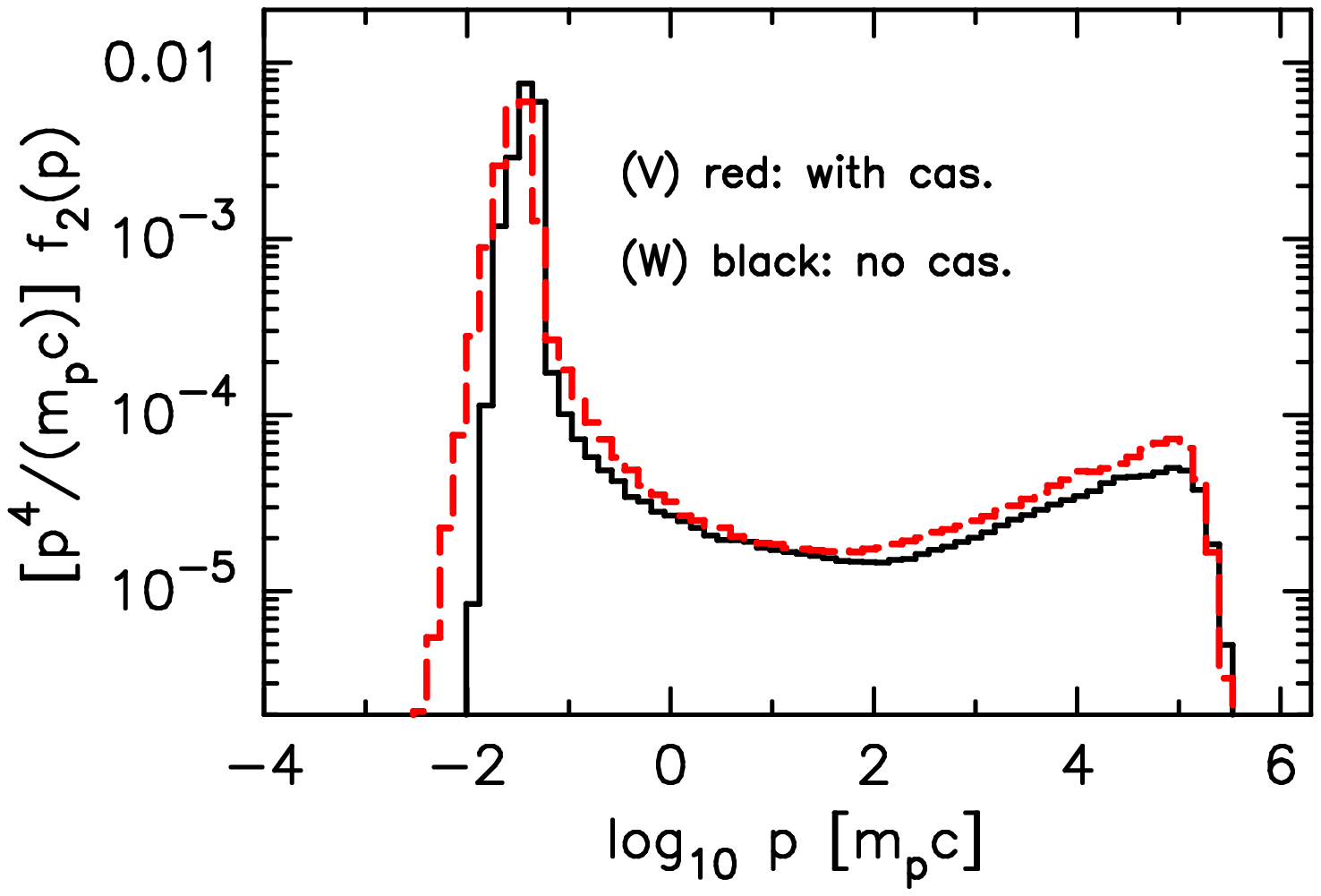}
\includegraphics[width=180pt,height=180pt]{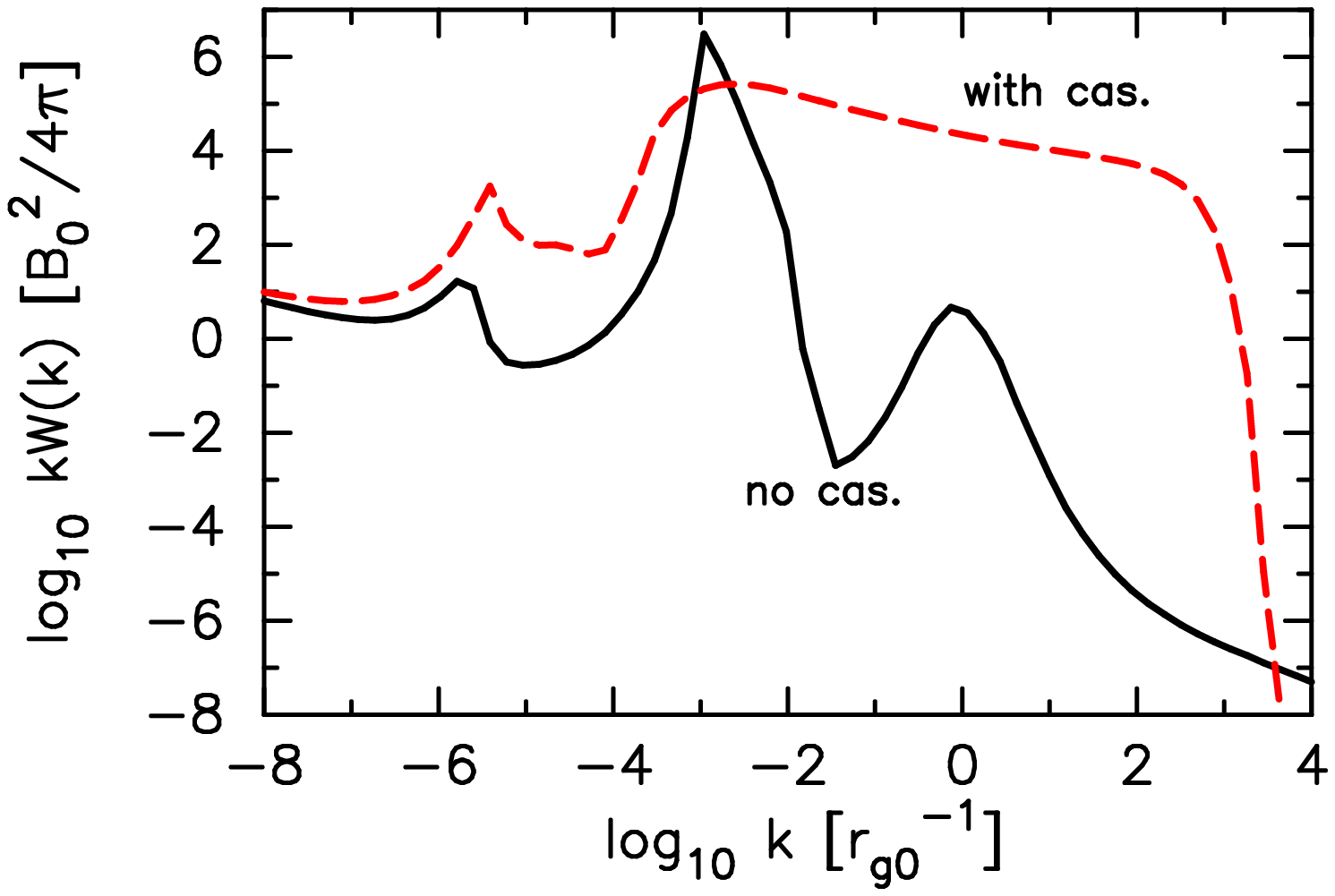}
\caption{{\it Left Panel:}  Particle momentum distribution in the shock downstream simulated with the non-linear Monte Carlo diffusive shock acceleration model by \citet{beov14} with (red curve) and without turbulent cascade (black curve). Shocks of 5,000 $\kms$ propagate in a plasma with initial number density $n$=0.3$ \cmc$ and temperature T~=~10$^6$~K. {\it Right Panel:} Simulated magnetic field fluctuation spectra in the downstream of the same shock as in the left panel. }
\label{fig:DSA:particle_field_spectra}
\end{figure}

A distinctive feature of collisionless  shocks in extended astrophysical flows is their ability to transfer a sizeable fraction of mechanical power of the flow to non-thermal particles and fluctuating magnetic fields by means of the first order Fermi  acceleration mechanism called the diffusive shock acceleration  \citep[see, e.g.,][]{bell78, Axford81, Drury83, be87, berkrym88, je91, MalDru01, bdd08, treumann09, bt11, schureea12,blasiAARv13}. The high efficiency of particle acceleration which may be well above 10\% as deduced, in particular, from observations of young supernova remnants \citep[see, e.g.,][]{vink12,helderea12,blasiAARv13} implies strong coupling between the accelerated particle population and the shock structure. The coupling is realized through the electromagnetic fluctuations carried with the shock flow and responsible for inelastic scattering the particles. The fluctuations are strongly amplified by CR driven instabilities  on a wide range of scales,  thus making the shock modeling an essentially nonlinear problem. Namely, the ponderomotive force acting on plasma due to the CR pressure gradient decelerates the upstream flow on a mesoscale which is much large than the CR gyroradii for non-relativistic shocks. Therefore the strong collisionless shock modeling is a complex multi-scale nonlinear problem. We shall not discuss here the details of diffusive shock acceleration models, referring the reader to numerous papers and reviews mentioned above, but rather illustrate in Figure~\ref{fig:DSA:particle_field_spectra} a simulated CR distribution and magnetic field spectra in the downstream of a shock of 5,000 $\kms$ in a plasma of temperature 10$^6$~K and number density 0.3~$\cmc$. The Monte Carlo simulations of non-linear diffusive shock acceleration were performed with a model account of turbulence amplification by CR-driven instabilities and some semi-phenomenological modeling of turbulence cascading. In strong shocks magnetic fluctuations are amplified by CR-driven instabilities and their magnetic spectra span a wide dynamical range of wavenumbers from macroscopic fluid scales down to the thermal ion and electron scales where the dissipation occurs. The spectra of particles and waves in the acceleration region simulated with Kolmogorov-like cascade models are shown in Figure~\ref{fig:DSA:particle_field_spectra} by red curves.

 It is important to note that one should always be careful and distinguish energetic particles confined within an accelerator and CRs that are escaping it because they may have drastically different spectra. The problem of CR escape from accelerators is fundamentally important for the connection of the observed  sources of non-thermal emission to the galactic pool of CRs and their global energetics. In popular models of diffusive shock acceleration particles mostly escape with energies close to  the highest  energy achieved at the given evolution stage, which is changing with time. This is a complicated non-linear problem some aspects of which were addressed in  \citet{je91}, \citet{pz05}, \citet{revillea09}, \citet{gabici_escape09}, \citet{Cap10}, \citet{eb11}, \citet{dtp12}, and \citet{bell13}. The fate of the rest of the confined low energy particles  which are subject of adiabatic deceleration in an expanding CR accelerator at the stage when shock acceleration is slow and which determine the eventual efficiency of CR injection into the ISM has not been studied in detail yet.

The fluctuating magnetic field in the shock downstream may be highly amplified by CR-driven instabilities. A strong non-relativistic shock may transfer up to 10\% of the ram pressure into magnetic fields as it is illustrated in the left panel of Figure~\ref{fig:DSA:MFA_T2}. There are convincing observational evidences for strong non-adiabatic amplification of magnetic fields in the vicinities of forward shocks of a number of young SNRs including Cas~A, Tycho, SN~1006, RXJ~1713.7-3946 and others obtained via arcsecond angular resolution imaging of synchrotron structures in these SNRs with {\sl Chandra} \citep[see, e.g.,][]{vink12, bamba_SNRs_05, uchiyama_ea07, reynoldsea_SSRv12, helderea12, sn1006_chandra14, Chandra_Highlites14}. The mechanisms of magnetic field amplification which are associated with CR-driven instabilities were thoroughly discussed and reviwed  in \citet{be87}, \citet{bell_lucek01}, \citet{MalDru01}, \citet{zweibel03}, \citet{bell04},  \citet{LP10}, \citet{schureea12}, \citet{blasiAARv13}, \citet{bbmo13}.

It should be noted that CR acceleration may strongly affect thermodynamical properties of a shock and namely, the gas compression ratio and the downstream ion temperature. The escaping CRs carry away some energy of the incoming flow thus increasing the gas compression and correspondingly reducing the downstream ion temperature.  This is illustrated in the right panel of Figure~\ref{fig:DSA:MFA_T2}. The simulations by \citet{beov14} also revealed that the gas compression ratio may be a non-monotonic function of shock velocity. One should have in mind that these results were obtained for plane parallel shocks and therefore a more realistic geometry may add a factor. Also, a rigorous treatment of turbulence cascading and dissipation, which will be achievable with future numerical approaches, would allow us to check and improve current Monte Carlo models. Apart from the most popular diffusive shock acceleration, there are some models which consider multiple interactions of CRs with a shock with non-diffusive test particle transport between the encounters \citep[see, e.g.,][]{kirk_ea_AA96, zimbardo_perri_Levy13}. The super-diffusive particle propagation may affect both the acceleration efficiency and the spectra of particles and magnetic turbulence. In the case of shocks propagating through partially ionized plasmas charge-exchange particle collisions provide a return flux of highly super-thermal neutrals heating the upstream plasma reducing the fluid Mach number and the compression ratio in Balmer-type shocks of velocities below 3,000 $\kms$ \citep[see, e.g.,][]{blasiea12}. The efficient CR acceleration may also modify the structure and the ultraviolet-optical-infrared  emission spectra of MHD radiative shocks \citep[see, e.g.,][]{BMRKV_SSRv13}.

\begin{figure}
\includegraphics[width=170pt]{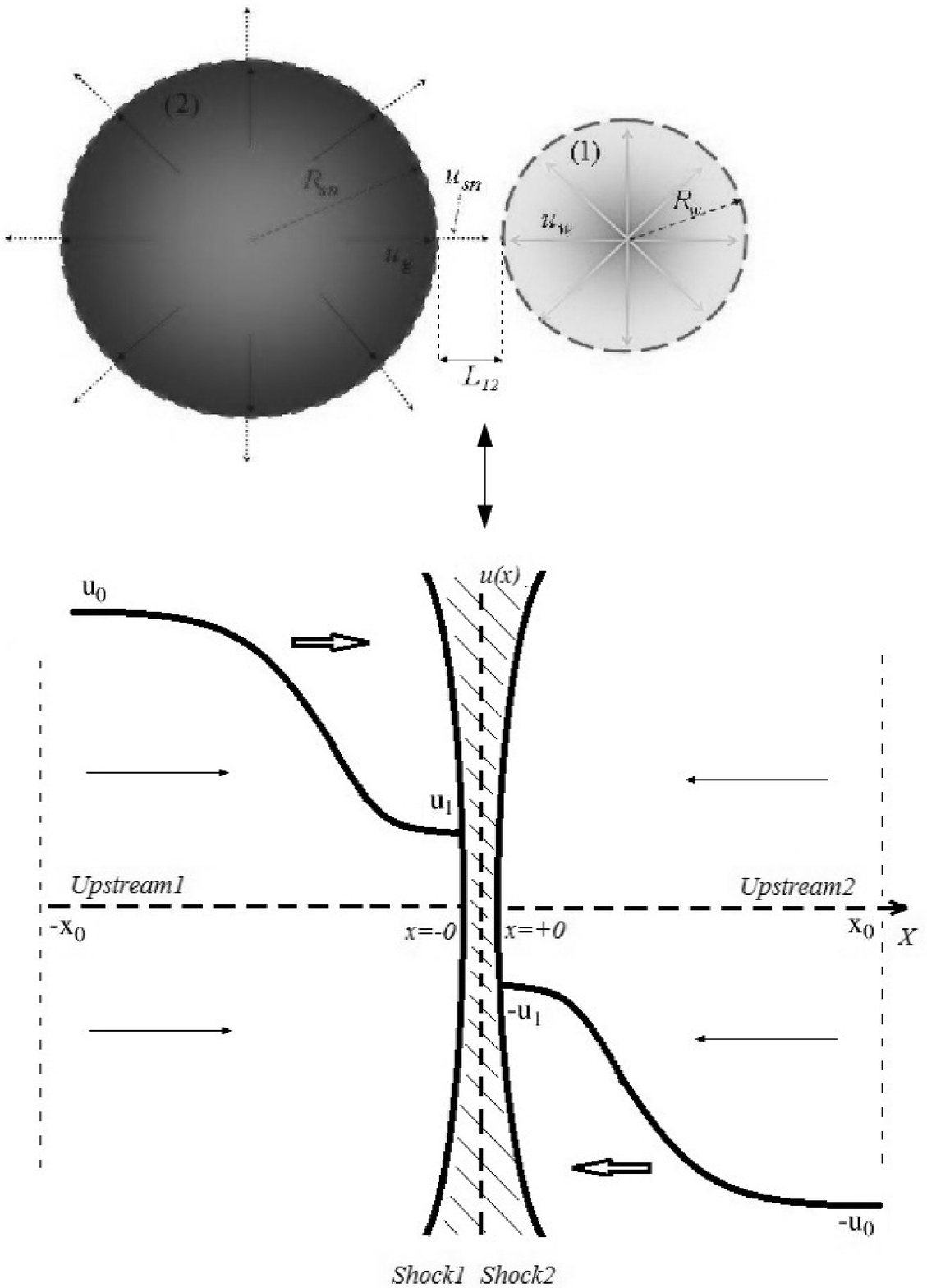}
\includegraphics[width=200pt]{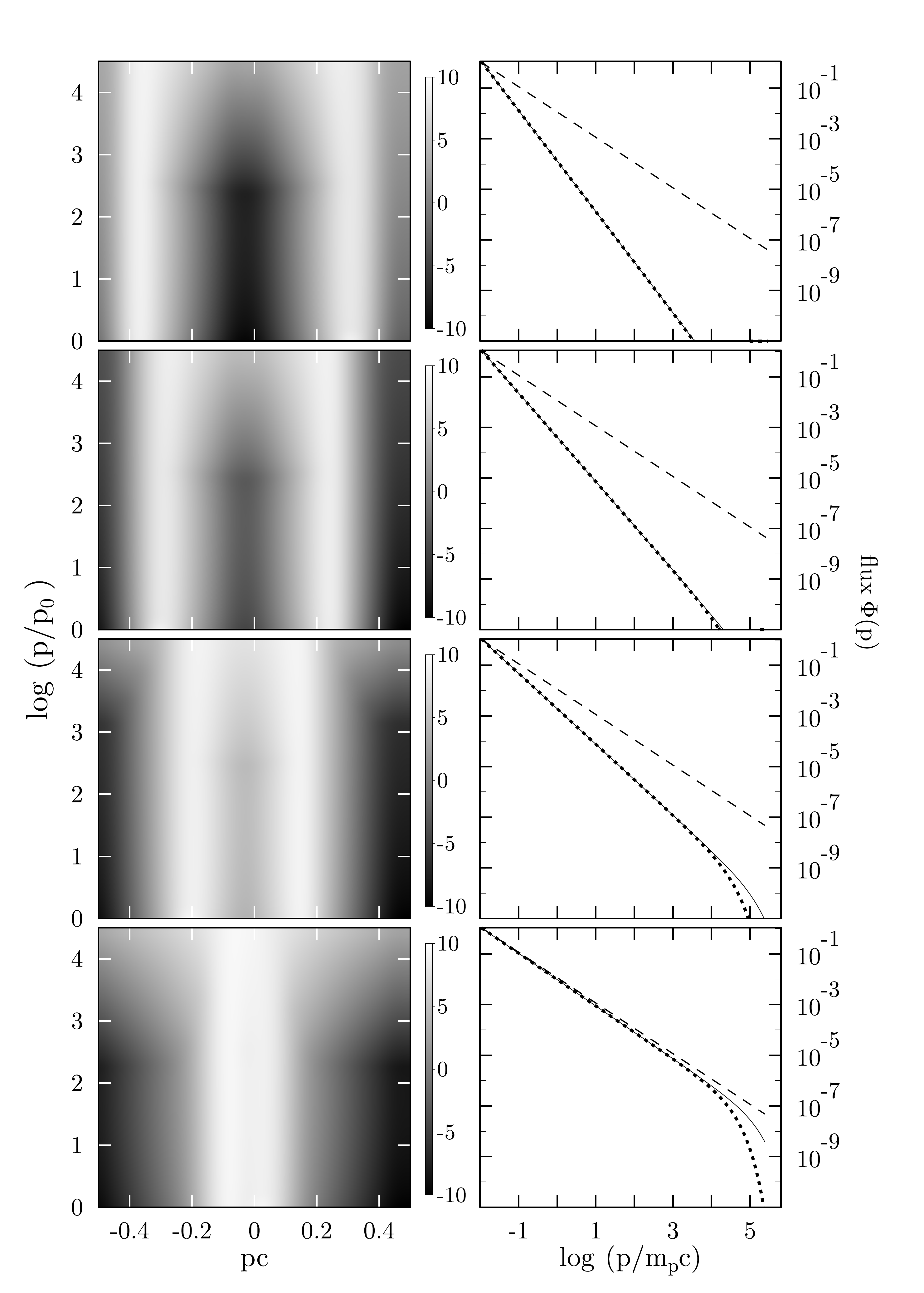}
\caption{{\it Left Panel:} The sketch shows an SNR shock (left circle) approaching  a termination shock of a wind of a young massive star (right circle). At the stage when the shocks are close enough providing the distance between the shocks L$_{12}$ is less then the shocks radii, the flow geometry may be approximated as plane parallel. High energy CR particles which have mean free paths larger than L$_{12}$ may cross the region between the shocks back and forth experiencing only head-on collisions with magnetic fluctuations in the approaching flows. Such particles are subject to Fermi~I acceleration. {\it Right Panel:} The spatial distribution (gray scale) and differential spectra $\Phi(p)= dN/dp(p)$ of the accelerated high energy particles taken at different evolution stages simulated with a non-linear kinetic model by \citet{MNRAS_BGO13}. The panels from top to bottom are corresponding to four different distances between the approaching shocks – 0.6, 0.5, 0.3, 0.1 pc respectively. The proton spectrum is shown by the solid curve while the electron spectrum affected by the synchrotron and Compton losses is shown by the short dashed curve. The dotted curve shows the expected form $\Phi(p)= dN/dp(p) \propto p^{-1}$ of the time asymptotic spectrum of the protons. } \label{fig:CSF1}
\end{figure}

We have discussed above observational evidences for the presence of non-thermal components in colliding stellar winds, supernovae in compact clusters of massive stars, and superbubbles. All of the systems  include ensembles of shocks of different strengths and likely contain strong MHD turbulence produced by fast supersonic flows in the sources of kinetic energy release.  More complex flows than just  isolated shocks  are expected in starburst regions. Therefore, we shall discuss specific features of particle acceleration in colliding shock flows and then acceleration by multiple shocks and large scale compressive turbulence.

\subsection{Particle acceleration in colliding shock flows: compact stellar clusters}\label{CSF}

The mechanism of CR acceleration via particle collisions with moving magnetic fields suggested by \citet{Fermi_PR49, Fermi_ApJ54} has a number of different realizations  and it is widely used in many astrophysical applications \citep[see, e.g.,][]{Axford81, top1985, jones94, biermann97, Pelletier01}. One of the most efficient versions of Fermi-type acceleration can be realized in the case of two colliding shocks driven by supersonic and super-alfvenic flows carrying magnetic fluctuations of different scales.  This case is illustrated in Figure~\ref{fig:CSF1} by a sketch of a forward supernova shell shock approaching a termination shock of a fast wind of a massive star. At the stage when the shocks are close enough providing the distance between the shocks L$_{12}$ is less then the shocks radii, the flow geometry may be approximated as plane parallel.  If the flows carry magnetic fluctuations, they scatter energetic particles providing some mean free path $\Lambda(p)$ which is usually energy-dependent.  The mean free path is likely larger inbetween the shocks than in a close vicinity of a shock. High energy CR particles which have mean free paths larger than L$_{12}$ may cross the region between the shocks back and forth experiencing only head-on collisions with magnetic fluctuations in the approaching flows. Such particles are subject to a very efficient first order Fermi acceleration.  The flow ram pressure can be efficiently converted into the pressure of energetic non-thermal particles both confined in the accelerator and escaping the source.

The pressure of high energy particles may modify the shock velocity profiles decelerating the flows as it  is shown in the bottom of the left panel of Figure~\ref{fig:CSF1} in the rest frame of the unperturbed ISM inbetween the shocks. On the other hand, the velocity profiles determine the spectra of accelerated particles. Therefore, the model is non-linear and the backreaction of the accelerated particles on the plasma flow is accounted for using in a semi-analytical way similar to that was developed by \citet[][]{Mal97}, \citet[][]{blasi02} and \citet[][]{AB05}. This model assumes a Bohm-like recipe of diffusive propagation of CR particles in the close vicinity of the shocks, supposing a high level of CR-driven instabilities amplified by magnetic fluctuations \citep[e.g.,][]{bell04, schureea12, beov14}, while the mean free path  of the highest energy particles  in between the shocks is larger than L$_{12}$. In this system particle distribution is nearly isotropic even for high energy particles with  $\Lambda(p) \gsim  L_{12}$ and the kinetic equation reduces to the so-called telegraph equation.

 The right panel of Fig.~\ref{fig:CSF1} shows the results of non-linear time-dependent modeling of energetic particle distribution in the vicinity of two closely approaching fast plane-parallel  MHD shocks at different distances between the shocks (from the top to bottom) as it was simulated by \citet{MNRAS_BGO13}. The gray scale panels in Fig.~\ref{fig:CSF1} illustrate the spatial distribution of particles of different energies at different moments of time. The proton spectra are approaching the asymptotic power-law distribution, which corresponds to a very hard spectrum $N(\gamma) \propto  \gamma^{-1}$  before the break (the dotted line in the far right spectral panels). The electron spectrum in this simulation is similar to that of proton, but has a maximal energy below that of the protons because of the synchrotron and inverse Compton energy losses. The electrons and positrons in this case are confined within the accelerator and only a fraction of high energy protons can escape the system. If the synchrotron or Compton losses are not strong enough  then the highest energies of electrons and protons are similar and electrons might escape the accelerator as well.

 The hard spectra of the accelerated particles confined in the acceleration zone result in very hard emission spectra with rather sharp breaks of  high energy radiation both of leptonic and hadronic origin. The modelled objects may appear as ``dark accelerators'' with a peaked spectral energy distribution of the high energy non-thermal radiation, but no apparent counterparts at longer wavelengths. The protons which escape the accelerator at different stages of the system evolution and produce pions via collisions with the gas in the surrounding dense shells or clouds  may radiate spectra of various hardness depending on the time evolution of their maximal energy. It is important to note that the spectra of escaped protons integrated over the lifetime of the accelerator are typically significantly softer than the spectra of particles confined within the source and may have spectral indexes close to -2. The exact value of the spectral index of the escaped particles which are contributing to the diffuse galactic CR population depends on the details of the complex evolution of the source.

In the case of a supernova shock of velocity  $u_s$ approaching a fast wind of a massive star of velocity $u_w$ in Bohm diffusion regime with $\Lambda(p) \approx R_g(p)$, the particle acceleration time $\tau _a$ can be estimated as
\begin{equation}\label{TauAc1}
\tau _a(\gamma)  \approx  \frac{c R_g(\gamma)}{u_s\,u_w},
\end{equation}
 where $R_g(\gamma)$ is the gyro-radius of a CR particle. The scale of the magnetic field amplified by CR-driven instabilities in this system can be approximated as $B \approx \sqrt{4\pi \eta_b \rho}\,u_s$, where the efficiency of magnetic field amplification $\eta_b \sim$ 0.1 (c.f. the left panel in Fig.~\ref{fig:DSA:MFA_T2}) and $\rho \approx m_p n$ is the ambient density. Then the acceleration time is $\tau _a \approx 2\,\cdot10^{10}\, \gamma_6 \, (\eta_b n)^{-0.5}\,u_{s3}^{-2}\,u_{w3}^{-1}$ s, where $u_{s3}$ and $u_{w3}$ are measured in the units of 10$^3\, \kms$ and CR proton Lorentz factor $\gamma_6$ is in the units of 10$^6$. The free expansion phase lasts about 250 yrs $(M/\Msun)^{5/6} n^{-1/2} E_{51}^{-1/2}$, where the velocity of an ejected mass $M$  is constant at about 10$^4 (M/\Msun)^{-1/2}E_{51}^{1/2}\, \kms$.

\subsection{Neutrinos and gamma-rays from compact
clusters}\label{neutrino_clusters}
In the case of a young supernova shock propagating through the winds of
massive stars in the compact cluster Westerlund~I where $n \sim$ 0.6 $\cmc$\
\citep{munoea06}, one can estimate $\tau _a \approx$ 400~yrs  for a proton of
energy about 40~PeV, if  $u_{s3} \sim$ 10 and  $u_{w3}\sim$ 3. The high
velocity of a supernova shock $u_{s3} \sim$ 10 is expected at the free
expansion stage if the mass of the fast ejecta is about of a few solar
masses. Therefore, the maximal energy of the protons accelerated at a few
hundred years old supernova shock approaching a fast wind of a massive star
may reach 20-40~PeV and thus, supernovae in compact clusters of massive stars
should be considered as potential "PeVatron" sources. It should be noted that
the estimations above are rather optimistic since the real geometry of
such a system is likely to
be more complicated.  However, if a supernova explosion occurs a few parsecs
outside of the core of a cluster of massive stars, the geometry factors
might be alleviated since the supernova shock may collide with the
cluster-scale wind of  $u_{w3} \gsim$ 3.

The efficient CR acceleration in the colliding shock flow lasts for about a
few hundred years when a substantial part of the mechanical power of
the supernova
shell is transferred to PeV energy CRs with a very hard spectrum. The
spectra of gamma-ray and neutrino emission from CRs inside the acceleration
zone have to be derived with an account for both the pion decay hadronic
emission and the inverse Compton components. The corresponding radiation
cross sections can be found in \citep[ e.g.,][]{aharonian_book04,
Kelner2006, Kamae06}. During this relatively short stage the accelerator can
be detected as a source of high energy radiation of both hadronic and
leptonic origin.  For the distant sources of  PeV emission  the gamma-ray
opacity of the extragalactic background light must be also accounted for
\citep[see, e.g.,][]{aharonian_book04, Dwekea13}. At the later stage the CRs
initially confined in the accelerator are released and the escaped particles
diffuse through the ambient ISM producing PeV regime photons and neutrinos
due to hadronic CR interactions.

\begin{figure}
\includegraphics[width=300pt,trim=50 390 20 50]{{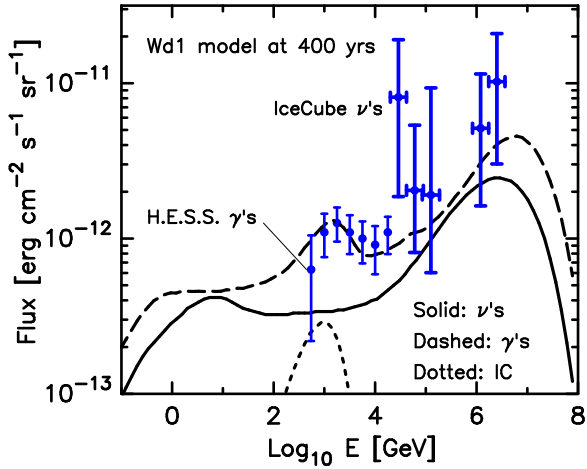}}
\caption{Spectra of the  neutrino and gamma-ray emission from a PeVatron source simulated in a model of particle acceleration by a free expanding shock of a supernova remnant colliding with a wind of a compact cluster of massive stars. Both the neutrino (solid line) and gamma-ray emission spectra (dashed line) are from the pion decays produced due to inelastic $pp$ collisions of protons accelerated by the short lived accelerator with the  surrounding clouds. The spectra of particles accelerated by colliding shocks may be significantly harder than that accelerated by young isolated supernova remnants. This results in the apparent PeV regime spectra upturn. For illustration purposes we put the source to the distance of the compact cluster Westerlund 1 and the data points are from H.E.S.S. observations of  Westerlund 1  by \citet[][]{hessWd1}. The data points for the H.E.S.S. source, and the five  Ice
Cube events explained in
Fig.~\ref{fig:NeuOnly}, are presented here just to illustrate how they compare to
the model predictions when the source is $\sim 400$\, yr old and
point-like, i.e., as we expect Westerlund I was about 10$^4$ years ago.
The simulated gamma-ray and neutrino emission at the present time from CRs
that escaped the accelerator in Westerlund I  $\sim 10^4$\, yr ago are
summarized in
Figs.~\ref{fig:gam1m} and \ref{fig:NeuOnly}.  The figure is adopted from
 "Ultrahard spectra of PeV neutrinos from supernovae in compact star clusters" (MNRAS v. 453,  p.113-121, 2015). } \label{Wd1_and_lines}
\end{figure}

In Fig.~\ref{Wd1_and_lines} we show the simulated spectra of neutrino and
gamma-ray emission produced by the CRs escaped from the accelerator and
diffusing into the surrounding cloudy medium. PeV regime particles are
expected to diffuse to distances  $\gsim$ 30 pc from the source in $\lsim$ 10,000 yrs after the release
from the acceleration site. The diffusion time roughly corresponds to the
estimated age of the supernova which has produced the magnetar CXOU~J1647-45
discussed in \S~\ref{Westerlund}, because the lifetime of the
accelerator is much shorter. A possible "PeVatron"  source at the distance
of Westerlund~I simulated with the nonlinear CR acceleration model of
\citet{MNRAS_BGO13} for the case of a supernova shock
of velocity $u_{s3} \sim$ 10 at the free expansion stage
colliding with a cluster scale wind of $u_{w3} \sim$ 3 as it was
discussed above.
The data points were taken from the H.E.S.S. observations by
\citep[][]{hessWd1}.
We do not discuss here the {\sl Fermi-LAT} source found in the
vicinity of Westerlund~I since
\citet{ohmea13} attributed the GeV emission to a pulsar wind nebula.
Supernovae interacting with fast winds of massive star clusters are
plausible sources of $\sim$ 40 PeV protons and PeV regime gamma-ray emission
and neutrinos, produced by the CRs in p-p collisions with shock-compressed
ambient medium. This can possibly explain some of the neutrino events detected recently by the {\sl Ice Cube Observatory} \citep{Aartsen13, ic_science13, Aartsen14} which are clumped within three
median angular errors from the direction to Westerlund I and the surrounding clouds associated with the TeV source detected by H.E.S.S.\footnote{After this review was published the model of Westerlund  I described above  was refined and the detailed version with comparison to 3 years Ice Cube data is published in
"Ultrahard spectra of PeV neutrinos from supernovae in compact star clusters" (MNRAS v. 453,  p.113-121, 2015). Therefore I  updated in this arxive version the Figures 10-12 which are adopted from the paper.}.

Diffusive shock acceleration in isolated SNRs is a likely mechanism to
accelerate the bulk of the galactic cosmic rays up to PeV regime energies
\citep{hillas05,bellea13,amato14}.  However, the type IIb supernovae, a subclass which
comprises likely about 3\% of the observed core collapsed SNe rate, was argued by
\citet{Ptu10} to be able to produce CRs of energies $\sim$ 100 PeV at the
early stages of SNR evolution. This type of sources may be in competition
with  SNR --- wind accelerators, which we have discussed above. The sources of
nonthermal emission and CRs in compact stellar clusters may contribute to
galactic CR spectrum in the PeV regime and above. Some spectral features
above the spectral knee indicated by recent observations \citep[see, e.g.,][]{iccube_lnA_APh13, berezhnevea13, KASCADE13} may be attributed to
individual contributions from some of the sources associated with the
compact stellar clusters.

Apart from a few parsec scale winds from the compact stellar clusters discussed above, a larger scale wind outflow of velocity $u_{w3} \lsim$ 0.3 was revealed in the Galactic Center region \citep[see, e.g.,][]{Bland-Hawthorn_GC_Wind03, GC_wind_multiwave_ApJ10, GC_cloud_goned_with_wind13, 2013Natur.493...66C}.  The large scale outflow from the Galactic Center region is likely connected to the recently discovered  "Fermi bubbles" \citet{2010ApJ...724.1044S} and to the two giant radio lobes \citep{2013Natur.493...66C}. Supernova remnants in the Galactic Center region may collide with the wind producing radio and gamma-ray sources with flat spectra which may account for some of the filamentary
non-thermal radio structures with flat spectra observed within the inner $\sim$ 100 pc vicinity of  the Galactic Center \citep[see, e.g.,][]{morris_serabyn96, YusefZadeh03}.

PeV neutrino sources with hard spectra  may contribute to starburst galaxy high energy radiation.  CR interactions in starburst galaxies are expected to  produce efficiently high energy gamma-rays and neutrinos.   \citet{loeb_waxman06} predicted that the cumulative contribution of starburst galaxies would produce a measurable neutrino background well above that is expected from atmospheric neutrino production.

\begin{figure}                            
\includegraphics[width=300pt,trim=50 390 20 50]{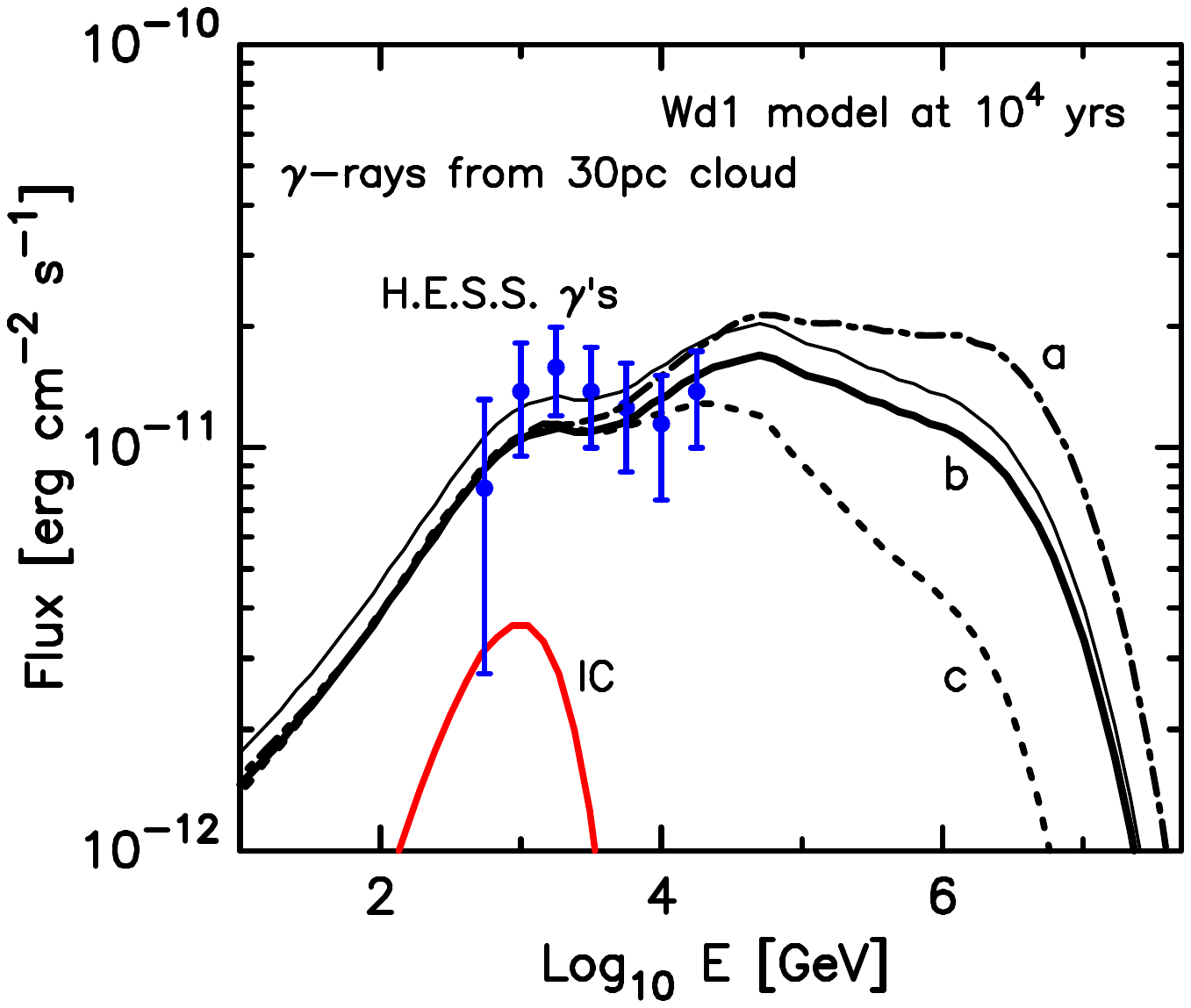}
\caption{Gamma-ray emission from inelastic $p-p$-interactions in the CSF
source at $\sim 10^4$\,yr after the SN explosion when CR protons produced
in the short-lived accelerator  have propagated into a nearby cloud
of $\sim 30$\,pc size. The magnetic field amplified by the CR-driven
instabilities in the vicinity of the fast shock in the CSF accelerator
were parameterized as  0.8 mG (c), 0.9 mG (b),
and 1 mG (a), all below 10\% of the ram pressure.
The IC curve is inverse Compton emission from the secondary electrons
produced by the inelastic $p-p$  interactions in the cloud. Only the
$\gamma$-rays from the H.E.S.S. field of view are included. The gas number
density of the nearby cloud is 25\,\pcc, except for the light-weight
solid curve where it is 30\,\pcc with $B=0.9$\,mG.  The figure is adopted from
 "Ultrahard spectra of PeV neutrinos from supernovae in compact star clusters" (MNRAS v. 453,  p.113-121, 2015). }\label{fig:gam1m}
\end{figure}

\begin{figure}               
\includegraphics[width=280pt,trim=50 140 0 250]{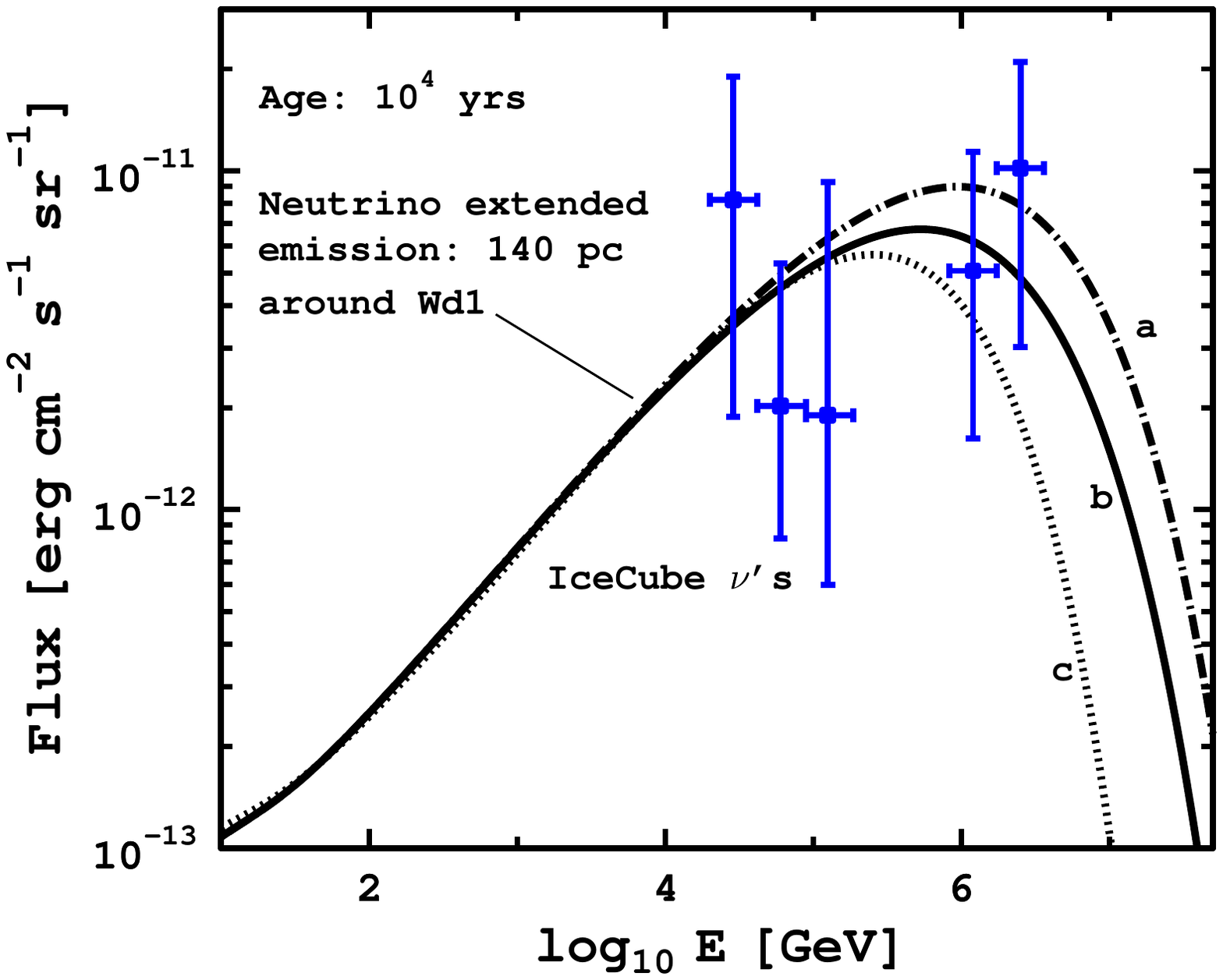}
\caption{Neutrino  emission from an extended ($\sim$ 140 pc radius)
source $\sim 10^4$\,yr after the SN explosion when CRs produced in the
short-lived accelerator  have propagated into the surrounding material.
The amplified magnetic fields in the short-lived acceleration source which determine
the maximal energy of accelerated protons are 0.8 mG (in the curve c), 0.9 mG (in the curve b),
and 1 mG (in the curve a). We note that the gamma-rays in Figs.~\ref{fig:gam1m} and the
neutrinos here originate from different volumes: only the $\gamma$-rays from
the H.E.S.S. field of view are shown in Fig.~\ref{fig:gam1m}, while the
neutrinos are from a larger region of radius 140 pc. The  neutrino data
points (1$\sigma$ energy flux error bars) are a subset of the
five events in the vicinity of Westerlund I  from  37 Ice Cube events of the first 3 years
\citep{Aartsen14}). The figure is adopted from
 "Ultrahard spectra of PeV neutrinos from supernovae in compact star clusters" (MNRAS v. 453,  p.113-121, 2015). }
\label{fig:NeuOnly}
\vspace{-1.\baselineskip}
\end{figure}

\subsection{Particle acceleration by long-wavelength turbulence and multiple shocks: superbubbles}\label{SBs}

Particle acceleration by individual SNRs and colliding shock flows may occur on time scales well bellow 1,000 yrs.
However, as discussed in \S~\ref{SBsHyd},  an OB star association on the time scales above a few hundred thousand years may enter the evolution stage with multiple SN explosions, which create large caverns of $\gsim$ 50 pc size --- galactic  superbubbles.
The injected mechanical power  may reach   $\sim 10^{38}\, \ergs$ over $\gsim$ 10 Myr --- the lifetime of a superbubble .
This process is accompanied by formation of multiple shocks, large scale flows and broad spectra of MHD fluctuations in
a tenuous plasma with frozen-in magnetic fields. The collective effect of multiple SNRs and strong winds of young massive stars in a superbubble may be able to re-energize CR particles  \citep[see][]{montmerle79, cesarsky_montmerle83, bf92, Axford94, higdonea98, klepach00, bt01, b01, ramaty_SB01, parizot04, marcowith05, Ferrand_Marcowith10} and even to extend the spectrum of accelerated particles to energies well beyond the Knee \citep[][]{bt01}.

 Nonthermal particles in superbubbles are scattered by small scale magnetic fluctuations
which determine their diffusive mean free paths $\Lambda(p)$, and also experience multiple interactions with shocks and large scale (i.e., of scales above $\Lambda(p)$) compression and rarefaction waves. Estimations of the mean free paths of super-thermal particles depend on rather uncertain hypothesis on the strength, polarization and statistical properties of small scale magnetic fluctuations which are produced either via cascading power from the stirring scales down to the scales of thermal gyroradii or by small scale instabilities of a local flow. Even in a single fluid MHD models cascading and dissipation in magnetic turbulence are rather complex processes \citep[see, e.g.,][]{K41, biskamp08, Schekochihin09, brandenburg11}, where a transition from collisional to collisionless plasma regimes occur, as it is the case in the starburst environment.

These processes are a lot more complex in the presence of cosmic rays. CR-driven instabilities strongly affect the turbulent dynamics at the mesoscopic gyrokinetic scales which span the range between the gyroradii of energetic particles and the scales of the energy-containing magnetohydrodynamic plasma \citep[see][and references therein]{bbmo13}. The CR-driven instabilities result in cross-scale coupling in the turbulence. In the lack of rigorous models of the magnetic turbulence one can rely on direct observations of collisionless turbulence in the Earth's magnetosphere and
in the heliosphere \citep{Matthaeus11, DdW13} where the anisotropic magnetic fluctuations with piece-wise power-law spectra are measured down to the electron gyroradius scale \citep{tu_Marsch95, Leamon98, 2005PhRvL..94u5002B, Alexandrovaea13}. Given the complexity of the multi-scale flows and  the lack of the observational data on the different scales, a practical way to treat particle acceleration processes in starburst regions and superbubbles on $\sim$ Myr time scale is to use coarse-grained kinetic equations which are consequently averaged over scales ranging from particle microscopic gyro-scales up to the macroscopic hydrodynamical scales determined by the repeated shock waves produced by supernovae and stellar winds. Within this approach the CR distribution function $F({\mbox{\boldmath $r$}},p,t)$ satisfies a kinetic equation subsequently averaged over an ensemble of  fluctuations of electric and magnetic fields induced by turbulent motions of highly conductive plasma with shocks of different strengths which introduce intermittency into the system.

\begin{figure}
\includegraphics[width=410pt]{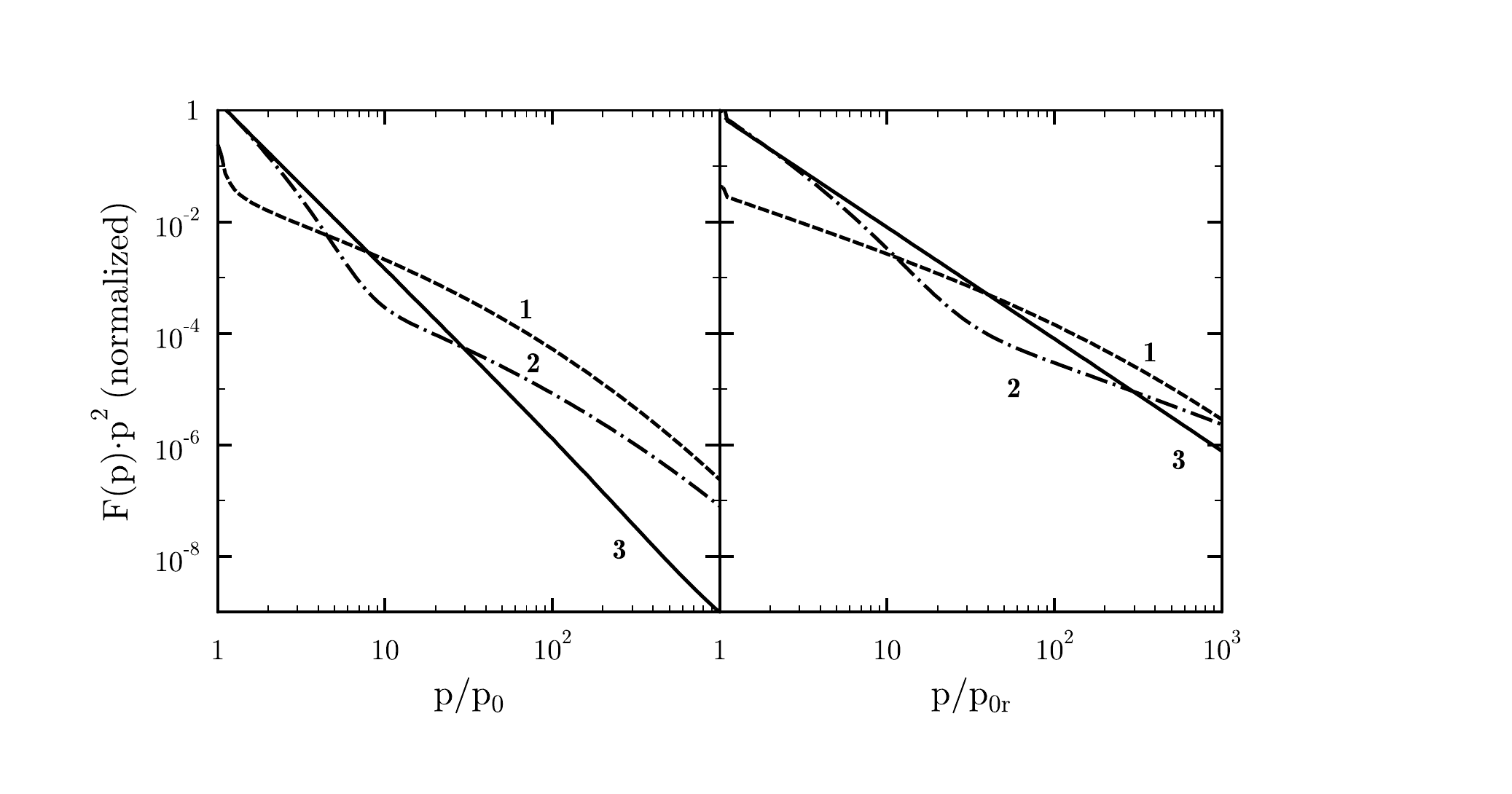}
\caption{ Temporal evolution of a particle distribution function in a superbubble simulated with the non-linear model of low-energy CR acceleration by \citet{b01}.
 Monoenergetic injection was assumed with the injection energies of about 10~keV (left panel) and 1 GeV (right panel). The turbulence energy-containing scale was 10 pc and
 the r.m.s. velocity amplitude was 1,000 $\kms$ for a superbubble of 50~pc size. The acceleration time was about 3$\times 10^5$ yrs. Curves 1, 2, and 3 show the CR spectra at 4, 6, and 10 acceleration times, correspondingly.} \label{time_evol}
\end{figure}

\begin{figure}
\includegraphics[width=430pt]{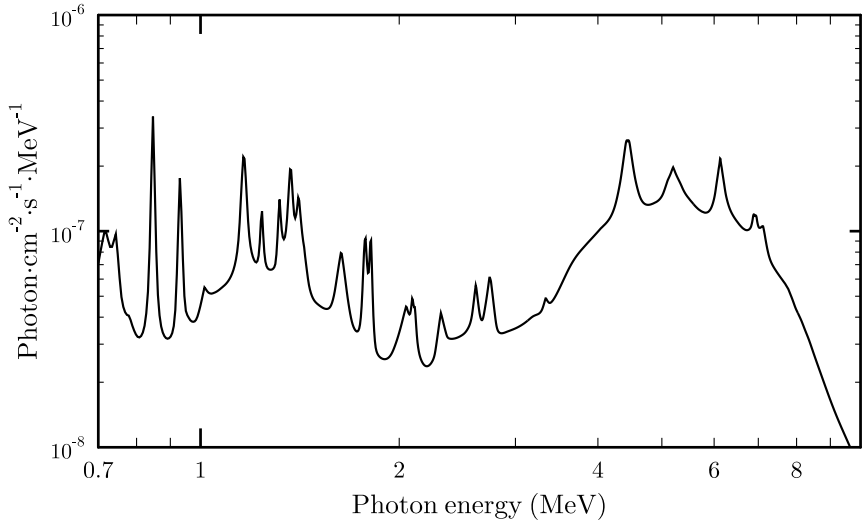}
\caption{A simulated spectrum of nuclear interaction gamma-ray line emission for a superbubble of a scale size of about 50 pc of the age $\sim$ Myr simulated with low energy CR particle spectra given in left panel  Fig.~\ref{time_evol}.} \label{lines}
\end{figure}

A statistical description of non-thermal particle distribution in intermittent systems with long wavelength compressible motions and multiple shocks was developed by \citet{bt94} who obtained a kinetic equation for the mean distribution function away from the strong shocks in the form
\begin{equation}
      \frac{\partial F}{\partial t} -
       \frac{\partial}{\partial r_{\alpha}} \: \chi_{\alpha \beta} \:
       \frac{\partial F}{\partial r_{\beta}}  =
       G  \hat{L} F +
      \frac{1}{p^2} \: \frac{\partial}{\partial p} \: p^4 D \:
      \frac{\partial F}{\partial p} + A {\hat{L}}^2 F +
      2B \hat{L} \hat{P} F + Q_{j}(p),
\end{equation}
the source term $ Q_{j}(p)$ is determined by injection
of the nuclei of a type $j$.
The integro-differential operators $\hat{L}$ and $\hat{P}$ are given by
\begin{equation}
      \hat{L}= \frac{1}{3p^2} \: \frac{\partial}{\partial p} \:
      p^{3-\delta} \: \int_{0}^{p} {\rm d}p' \: {p'}^\delta
      \frac{\partial}{\partial p'} \;;~~~~~
      \hat{P}= \frac{p }{3} \: \frac{\partial}{\partial p} \, .
\end{equation}
The index $\delta$ of the CR momentum distribution at a strong shock in the test particle case is about 4, but in non-linear models it may differ from this universal value.

The information on the turbulence and shocks is in the  kinetic coefficients $A$, $B$, $D$, $G$, $\tau_{sh}$, and
diffusion tensor $\chi_{\alpha \beta}$ which were expressed in terms of spectral functions describing correlations between shocks and the long wavelength (i.e., of scales $l\gg \Lambda(p)$) turbulent motions of compressible plasma. Short scale magnetic fluctuations determine the local CR scattering rate and therefore, the mean free path  $\Lambda(p)$ and its momentum dependence. The kinetic coefficients satisfy some transcendental renormalization equations given in \citet{bt94}and \citet{b01} which connect these to
the transverse and longitudinal parts of the Fourier components of the turbulent velocity correlation tensor and the
correlations between velocity jumps on shocks and also between shocks and smooth rarefactions.

Two distinct regimes of CR acceleration in a turbulent bubble characterized by an energy-containing scale $l$ with the amplitude of the bulk plasma speed $u$ should be recognized. At low enough CR energies of $p < p_{\ast}$,  where $\Lambda(p_{\ast}) =  l u/v$, the CR transport is determined by turbulent advection (here $v$ is the CR particle velocity). The diffusion tensor $\chi_{\alpha \beta}$ is nearly momentum independent. CR particles are strongly coupled to plasma motions, CR acceleration time is close to the turbulence turnaround time and particles may strongly affect the the supersonic turbulence. In a nonlinear model by \citet{b01} where a simplified description of the long-wavelength turbulence was considered, an efficient conversion of the shock turbulence power into CRs of momenta $p < p_{\ast}$ and a strong time evolution were found for particles injected initially with the same momentum $p_0 < p_{\ast}$. In nonlinear models no superposition principle holds and therefore,  the distribution function calculated for monoenergetic injection can not be used as a Green function.

In Figure~\ref{time_evol} the evolution of CR distribution is illustrated  with two models. In the first model shown in the left panel the proton injection momentum $p_0$ was chosen to be non-relativistic (a few keV). The CR spectra evolution in this nonlinear model follow soft-hard-soft scenario with time asymptotic power-law momentum distribution of non-relativistic particles $p^2F(p)  \propto p^{-3}$ (curve 3 in the left panel in Fig.~\ref{time_evol}).  After an appropriate scaling this case can be confronted against satellite observations of low energy CR population  in the corotation interaction regions of the heliosphere, where multiple shocks and compressive waves accelerate particles.

Analysing quiet epoch suprathermal ion distributions in the heliosphere \citet[][]{fisk_gloecker06, fisk_gloecker12}  pointed out that the observed suprathermal tails follow power laws $\propto v^{-5}$ and  the spectral slope of index "-5" is universal in this objects. The quiet epoch spectra should correspond to our mean distribution function, while in shock vicinities the spectra are harder. The universal index "-5" exactly corresponds to asymptotic power-law $p^2F(p)  \propto p^{-3}$ in
 Fig.~\ref{time_evol}. Long wavelength compressible motions are likely to play an important role in shaping CR  distributions in the interplanetary medium \citep[][]{fisk_gloecker12, JL10, zhang_lee13, asz13}. The low energy CR spectra in a superbubble could be tested with gamma-ray line observations in MeV regime. In the right panel in Figure~\ref{lines} we show a simulated spectrum of gamma-ray line emission which could be detected from a superbubble at a few kpc distance with  MeV regime detectors of sensitivity $\sim 10^{-7} \lfl$ which are not  available yet.

In the case of relativistic CR injection, but still in the regime of low energy CRs, where $p < p_{\ast}$, the proton spectrum in Fig.~\ref{time_evol}) again demonstrates a soft-hard-soft  behavior, but the asymptotic power-law distribution is  $p^2F(p)  \propto p^{-2}$ which is harder than that in the non-relativistic case. The pion production in $pp$ collisions of the protons in the energy range between GeV and TeV regimes accelerated in a superbubble may explain the  GeV-TeV  emission from  the Cygnus cocoon discovered with {\sl Fermi-LAT} by \citet{fermiSB11} and shown in Fig.~\ref{fig:Fermi_Cygnus}  \citep[see also][]{Cygnus_cocoon_ARGO-YBJ14}. Somewhat softer observed gamma-ray spectra  may be attributed to the elongated shape of the cocoon, as only a spherical superbubble was modelled. The elongated shape would alleviate CR escape from the cocoon resulting in softer spectra. The value of the maximal momentum $p_{\ast}$ depends on the amplitude of small scale collisionless turbulence which can not be resolved and can be estimated only indirectly, e.g., from electron density fluctuations derived from the observed interstellar scintillation of pulsar signals \citep[see, e.g.,][]{1995ApJ...443..209A}. One can expect the transition energy  $cp_{\ast}$ to be at  TeV regime in an extended superbubble. A possible presence of a 'kink' in CR particle spectra at about 0.2 TeV per nucleon  found in direct measurements of galactic CRs was attributed by \citet{Erlikin_Wolfendale12} to propagation and acceleration processes
in the Local Bubble.

\begin{figure}
\includegraphics[width=190pt]{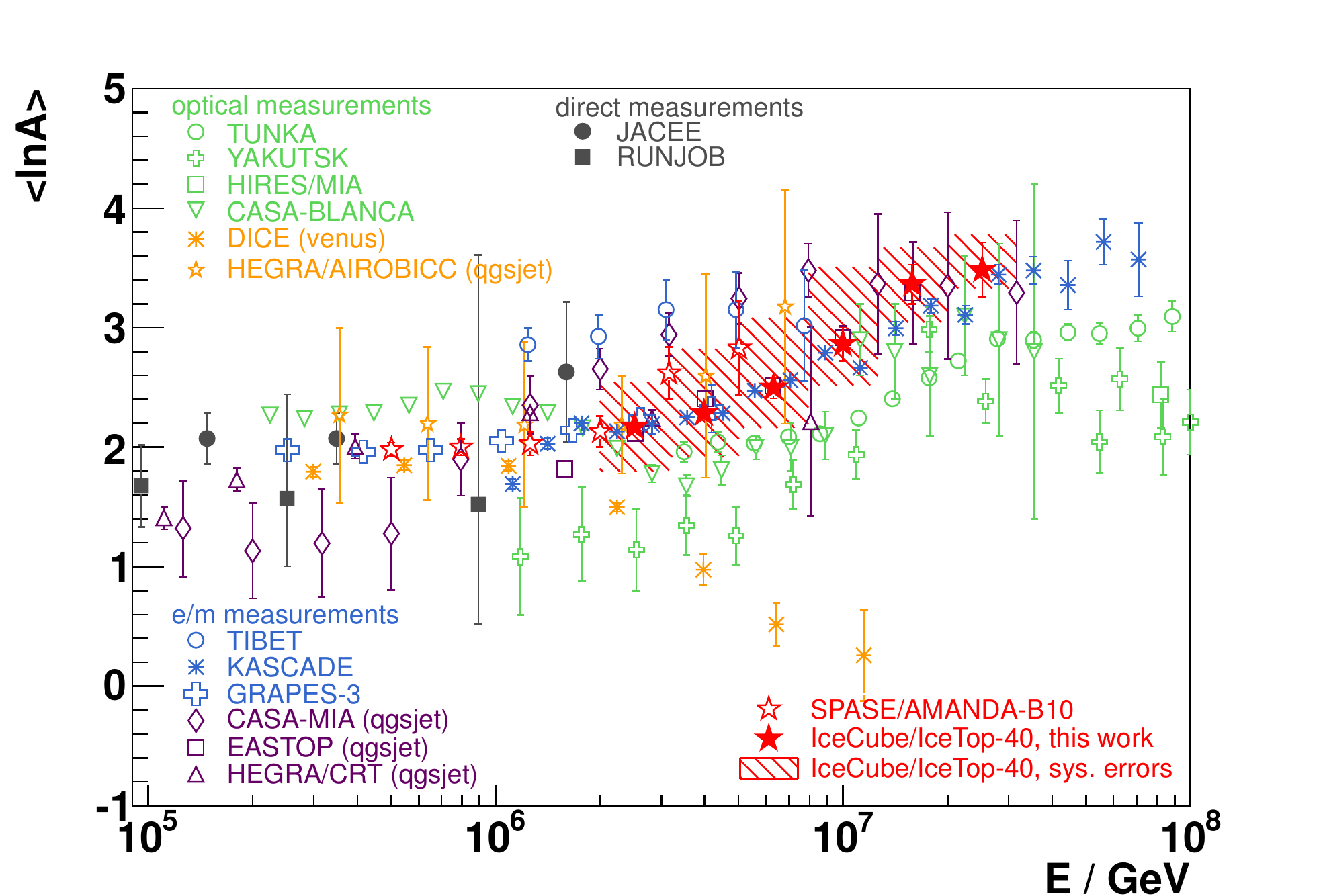}
\includegraphics[width=170pt]{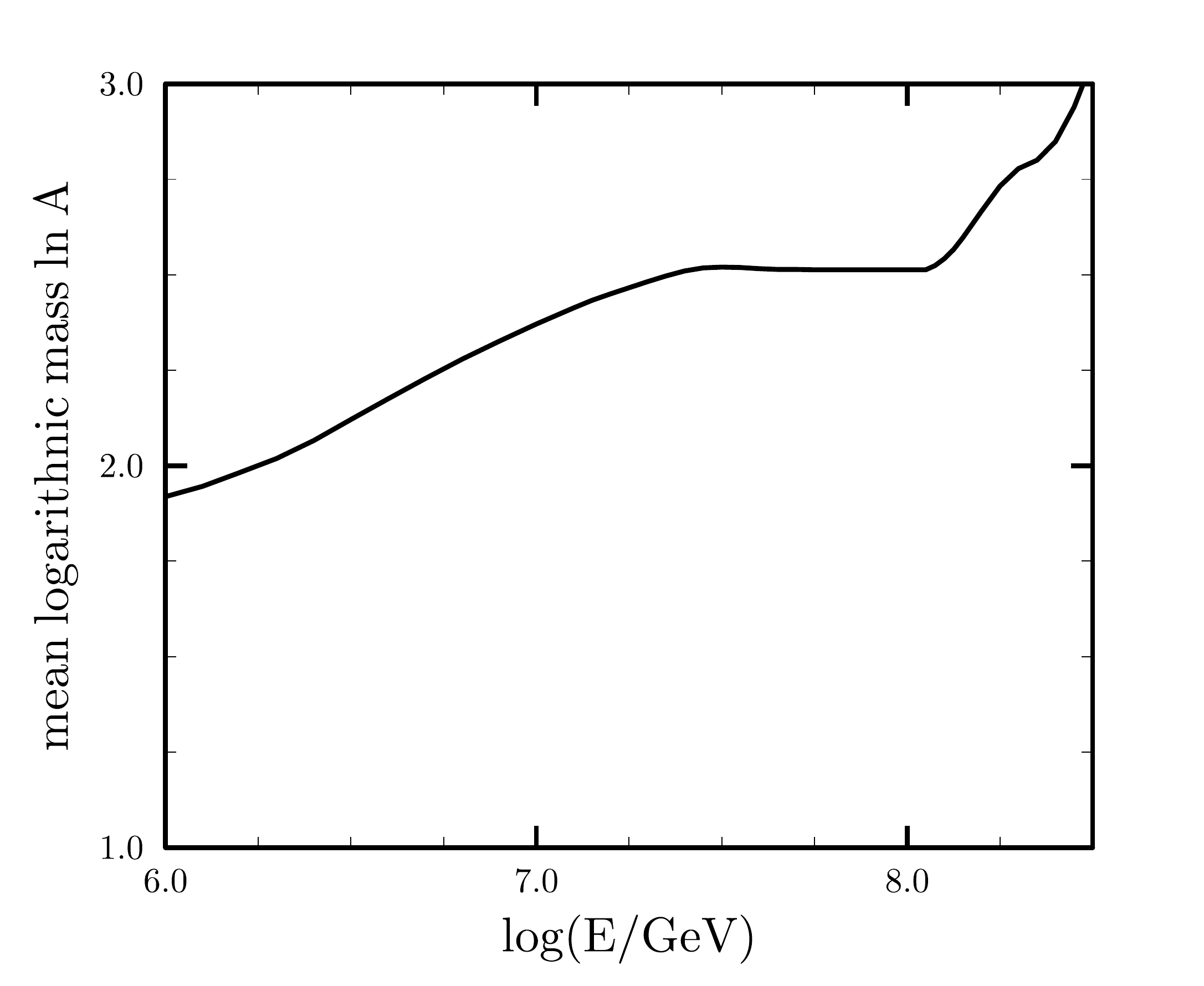}
\caption{{\it Left Panel:} The mean logarithm mass of galactic cosmic rays as observed by different detectors and compiled by \citet[][]{iccube_lnA_APh13}. {\it Right Panel:} The mean logarithm mass of CRs accelerated in a superbubble as simulated by \citet{bt01}. }
\label{fig:CR_superbubble}
\end{figure}

High energy ions of $p > p_{\ast}$ accelerated in superbubbles may contribute to observed distribution of galactic cosmic rays even at energies above the Knee where a substantial enrichment with heavy ions is expected in the superbubble model. In the right panel of Figure~\ref{fig:CR_superbubble} we present the mean atomic weight of CRs as a function of CR energy as it was simulated by \citet{bt01}. This can be compared to the recent compilation of observations available from \citet[][]{iccube_lnA_APh13}. Models of CR acceleration in superbubbles were argued by \citet{higdon_ling05} as a favourable way to explain the anomalous $^{22}$Ne/$^{20}$Ne ratio in CRs (c.f. \citet{casse_paul82}), the cosmic-ray actinide/Pt group and UPuCm/Th ratios, and the constant LiBeB/(C+O) ratio observed in very old, metal-poor stars \citep[see also][]{parizot04, binns_SB08}. However, \citet{prantzos12} has recently argued against the superbubble model of  the $^{22}$Ne/$^{20}$Ne CR isotope anomaly, thus further investigation of the superbubble models of CR origin is needed.

\section{Concluding remarks}
A significant progress has been recently made in detection and
investigation of gamma-ray spectra of starburst galaxies M82, NGC~253,
NGC~1068 with the {\sl Large Area Telescope} aboard {\sl Fermi
gamma-ray observatory} and imaging atmospheric
Cherenkov telescopes H.E.S.S., MAGIC, and VERITAS. The observations
revealed the presence of efficient mechanisms
of particle acceleration at least up to the TeV range. For the first
time gamma-ray emission was detected from the extended
cocoon associated with OB-star complexes in Cygnus X, which may
represent the class of galactic superbubbles.
High resolution X-ray imaging and spectroscopy of compact clusters of
massive stars  Westerlund~I, Arches,  Quintuplet and the superbubbles in LMC with {\sl
Chandra}, {\sl XMM-Newton}, {\sl INTEGRAL}, and {\sl NuSTAR} X-ray
observatories provided clear evidences of multi-component emission
which is produced by thermal and non-thermal sources due to violent
processes of wind collisions and to compact remnants of supernovae.

In general, processes of particle acceleration in the violent
starburst regions are due to efficient conversion of kinetic power of
the fast outflows of winds from massive stars and supernovae into
energetic particles and fluctuating magnetic fields, which occurs in the
vicinities of fast shocks and colliding supersonic flows.
The efficiency of kinetic power conversion into cosmic rays may reach
tens of percent which indicates a substantial backreaction
of the accelerated particles on the flow dynamics.
Therefore, MHD modeling of plasma heating and large scale
wind formation in compact clusters of massive stars and superbubbles
should account for the pressure of non-thermal particles
and magnetic fields.

Nonlinear models of energetic particle acceleration
discussed in this review illustrate the diversity of the acceleration process.
Supernova shells colliding with a cluster wind may give birth to
compact sources with hard spectra  of PeV regime CRs accelerated on a time scale shorter than 1,000 yrs.
The subsequent escape of the accelerated particles should produce extended sources of gamma-rays and neutrinos.
  This can be tested with the next generation of Cherenkov telescopes
and with the existing neutrino telescopes like the South Pole neutrino observatory {\sl Ice Cube}.

CRs may be efficiently accelerated and confined in superbubbles on Myrs time scale.
The long lived superbubbles may contain a low-energy CR component,
whose partial pressure is substantial.
The low-energy component can be constrained via a search for MeV range gamma-ray lines,
which would require a new generation of sensitive
spectrometers to appear.
Being an energetically significant the long-lived CR component produced during collisionless
relaxation of violent plasma outflows in starburst environments provide
ionization of the starforming clouds and thus it is an important integral part of galactic ecology.

\begin{acknowledgements}
The invitation to this review by T.~Courvoisier is deeply acknowledged. The author benefited from his collaborations with A.~Artemyev, C.~Georgy, I.~Grenier, D.C.~Ellison, G.~Fleishman, P.~Gladilin, M.~Gustov, A.~Marcowith, K.P.~Levenfish, D.~Folini,  S.~Osipov, E.~Parizot, Yu.~Uvarov, R.~Walder. Special thanks are due to I.N.~Toptygin and A.M.~Krassilchtchikov for their help at various stages of my work.
\end{acknowledgements}


\end{document}